\documentclass[sn-mathphys,Numbered]{sn-jnl}

\usepackage{graphicx,subcaption}%
\usepackage{adjustbox}
\usepackage{multirow}%
\usepackage{amsmath,amssymb,amsfonts}%
\usepackage{amsthm}%
\usepackage{mathrsfs}%
\usepackage[title]{appendix}%
\usepackage{xcolor}%
\usepackage{textcomp}%
\usepackage{manyfoot}%
\usepackage{booktabs}%
\usepackage{algorithm}%
\usepackage{algorithmicx}%
\usepackage{algpseudocode}%
\usepackage{listings}%
\usepackage[section]{placeins}
\usepackage{makecell}

\theoremstyle{thmstyleone}%
\theoremstyle{thmstyletwo}%

\theoremstyle{thmstylethree}%

\begin{document}

\title[WIDESim]{WIDESim: A toolkit for simulating resource management techniques of scientific \underline{W}orkflows \underline{i}n \underline{D}istributed \underline{E}nvironments with graph topology}


\author[1]{\fnm{Mohammad Amin} \sur{Rayej}}\email{m.amin.rayej@gmail.com}

\author[1]{\fnm{Hajar} \sur{Siar}}\email{hajar.siar87@sharif.edu}

\author[1]{\fnm{Ahmadreza} \sur{Hamzei}}\email{ahmadreza.hamzei@sharif.edu}
\equalcont{These authors contributed equally to this work.}

\author[1]{\fnm{Mohammad Sadegh} \sur{Majidi Yazdi}}\email{m.majidi@sharif.edu}
\equalcont{These authors contributed equally to this work.}

\author[1]{\fnm{Parsa} \sur{Mohammadian}}\email{parsa.mohammadian@sharif.edu}
\equalcont{These authors contributed equally to this work.}

\author*[1]{\fnm{Mohammad} \sur{Izadi}}\email{izadi@sharif.edu}

\affil[1]{\orgdiv{Department of Computer Engineering}, \orgname{Sharif University of Technology}, \orgaddress{\city{Tehran}, \country{Iran}}}


\abstract{IoT devices trigger real-time applications by receiving data from their vicinity. Modeling these applications in the form of workflows enables automating their procedure, especially for the business and industry. Depending on the features of the applications, they can be modeled in different forms, including single workflow, multiple workflows, and workflow ensembles. Since the whole data must be sent to the cloud servers for processing and storage, cloud computing has many challenges for executing real-time applications, such as bandwidth limitation, delay, and privacy. Edge paradigms are introduced to address the challenges of cloud computing in executing IoT applications. Executing IoT applications using device-to-device (D2D) communications in edge paradigms requiring direct communication between devices in a network with a graph topology. While there is no simulator supporting simulating workflow-based applications and device-to-device communication, this paper introduces a toolkit for simulating resource management of scientific workflows in distributed environments with graph topology called WIDESim\footnote{The simulation toolkit (WIDESim) can be downloaded from \href{https://github.com/ARH80/WIDESim}{\texttt{our Github repository}}}. The graph topology of WIDESim enables D2D communications in edge paradigms. WIDESim can work with all three different structures of scientific workflows: single, multiple workflows, and workflow ensembles. It has no constraint on the topology of the distributed environment. Also, unlike most existing network simulators, this simulator enables dynamic resource management and scheduling. We have validated the performance of WIDESim in comparison to standard simulators and workflow management tools. Also, we have evaluated its performance in different scenarios of distributed computing systems using different types of workflow-based applications. The results indicate that WIDESim's performance is close to existing standard simulators besides its improvements. Additionally, the findings demonstrate the satisfactory performance of the extended features incorporated within WIDESim.}




\maketitle

\section{Introduction}\label{intro}

Internet of Things (IoT) is a game-changer in the industry that brings various opportunities for different applications such as healthcare, agriculture, and transportation using connected devices to the internet. IoT devices gather information from their surrounding environment and send IoT data streams to the computation resources. Most of the IoT applications are large-scale and modeled as scientific workflows that are directed acyclic graphs (DAG) of interdependent tasks \cite{nardelli2017osmotic}. Workflows introduce a formal representation of the application, automating its process. According to the literature, there are three forms of scientific workflow applications: individual or multiple workflows and workflow ensembles. There are some differences between workflow ensembles and multiple workflows. The QoS requirements are determined for the ensemble, not each workflow. So, all workflows cannot be complete necessarily, and an additive objective compared with multiple workflows is maximizing the number of completed workflows from the ensemble. Also, the ensemble's workflows have a similar structure, and the differences are in the input data and workflows' size. The final difference is the number of workflows in the ensemble, which is predefined in most cases \cite{siar2021offloading, rodriguez2018scheduling}.\\
IoT services are commonly provided using cloud computing. However, this computing paradigm has a centralized model, transmitting all data to the cloud data-centers for processing and storing \cite{siar2021offloading}. Cloud computing providing centralized and near to infinite computation power is an appropriate infrastructure to process large-scale computation problems. However, an essential characteristic of IoT applications is the distribution of IoT devices and their large geographic distance from computing servers, also most IoT applications are delay-sensitive. Because of the slow growth of bandwidth than computing power, cloud computing tackles bandwidth delay, especially for bandwidth-hungry IoT applications such as video processing, virtual reality, and object detection. Also, sending all raw data to cloud servers owned by un-trustable providers brings privacy issues for data owners \cite{de2019foundations, siar2021offloading, karagiannis2021context,rabay2019fog}.\\ 
Fog computing and edge computing are introduced to advance the centralized nature of cloud computing architecture in executing IoT applications with various resource requirements. Fog computing has a hierarchical architecture of processing and networking resources in the spectrum of end devices to cloud devices. Edge/fog devices with geo-distributed and heterogeneous resources can process input data, so there is no need to send the raw data to cloud servers. This infrastructure has the potential to tackle the challenges of cloud-centric IoT. Architectures of distributed computing systems that use the processing, storage, and networking resources in the continuum from end devices to cloud are known as edge paradigms \cite{yousefpour2019all}, these architectures have various communication models between devices at all three layers: end devices, edge devices, and the cloud. According to the importance of device-to-device communication for exchanging information between devices, using this technology they can communicate with devices in the same layer or different layers. This yields a graph topology for the edge paradigms \cite{yousefpour2019all, tocze2018taxonomy}. \\
Resource management techniques must be used to unleash the potential of edge paradigms and optimal utilization of resources \cite{buyya2019fog}. These techniques introduce solutions for optimal execution of applications in the network \cite{tocze2018taxonomy}. There are several works in the literature for managing the execution of scientific applications in fog environments \cite{siar2021offloading, yu2015cooperative, stavrinides2019hybrid, chen2017socially, lim2020incentive, bianzino2014green, goudarzi2020application}. In these studies, various topologies are considered for the fog environments, from a tree to a fully connected, and different resource management problems are contemplated, such as scheduling, computation offloading, resource discovery, resource sharing \cite{tocze2018taxonomy}. An IoT use case can have a particular topology for the network structure. Hence the resource management solution is introduced for that topology. Also, because of the distributed infrastructure of edge/fog computing, researchers are introducing distributed, and decentralized algorithms for these environments \cite{zhang2019dmra, tripathi2017non, jovsilo2018decentralized, guerrero2019lightweight}. Therefore, resource management schemes use different network topologies and different algorithms depending on the type of application and system model. \\
Analyzing the performance of a resource management solution requires evaluation environments. Since the evaluation of solutions in real environments is costly and does not provide a repeatable and controllable condition, simulation tools are considered as a testbed in most studies \cite{gupta2017ifogsim}. An appropriate simulation environment for resource management techniques must be flexible in implementing the desired system model to simulate the solution's performance close to the real environment. Although by increasing the number of IoT devices, the use of the application in the form of scientific workflows will increase, there are limited tools for simulating resource management and scheduling of these applications in distributed environments. Most of the existing tools have limitations in terms of network topology and computational model, such that they can not work with all types of scientific workflows \cite{chen2012workflowsim, liu2019fogworkflowsim}. Also, there is no simulation toolkit supporting graph network topology to simulate decentralized resource management algorithms. On the other hand, the existing standard simulation tools do not support dynamic arrival of tasks, and hence dynamic resource management and scheduling.\\
This paper introduces a flexible toolkit called WIDESim for simulating resource management of workflows in distributed environments. WIDESim is an extension of the CloudSim simulator \cite{calheiros2011cloudsim} with the ability to simulate distributed environments of cloud, edge, and fog computing without any restrictions on the network topology. To the best of our knowledge, WIDESim is the first network simulator that is flexible in defining the topology of the distributed environments to consider both centralized and decentralized resource management schemes. On the other hand, WIDESim's computation model supports applications in all three different forms of scientific workflows.\\
The structure of this paper summarizes as follows: section \ref{rel} presents the related works of this study by mentioning related resource management and scheduling approaches in distributed computing systems, and the most popular and state-of-the-art network simulators. section \ref{arch} presents the architecture of the WIDESim simulator. Evaluation of WIDESim's validity is present in section \ref{eval}. Finally, section \ref{conc}, concludes the paper.

\section{Related Works}\label{rel}
The related works of this study can be divided into two sections, the first is resource management  approaches in distributed computing systems and the second is simulation tools for these approaches. 

\subsection{Resource management and scheduling}\label{resmanage}
Since resource management and scheduling are vital challenges in distributed computing systems, there are many studies on these issues \cite{hameed2016survey,zhang2010cloud,singh2016survey, yi2015survey, zhang2017hierarchical,mahmud2020application, xu2021privacy,li2020noma,goudarzi2020application, ijaz2021energy,siar2021offloading}. Among them, many studies considered the network topology as a graph to enable communication between devices in each layer. In \cite{xu2021privacy}, an edge device desire to offload its computation to one or more providers, while providers desire optimal pricing for their services. The nodes in the network are divided into two groups of leader and follower and there is a communication link between each leader and follower. The paper of \cite{li2020noma} presented a reinforcement learning-based approach for computation offloading in edge computing. There is one edge server in the system model and a group of edge devices communicates with each other to decide how to offload their computation on the edge server. The paper of \cite{hong2019multi} presented a game-theoretic approach for multi-hop computation-offloading in an edge–cloud computing to achieve the quality of service (QoS) in a distributed manner. The presented solution is using communication between devices for determining their computation offloading strategies. A scheduling solution is presented in \cite{zeng2016joint}, where devices must communicate with each other.\\
Managing an application with dependent tasks in the form of workflows requires considering the dependencies among tasks. A child task is ready provided that the execution of its parents is completed and their dependency data sent to the resource node that the child is assigned. Since the parent and child tasks can place in different layers of the edge/fog computing environment, communication between nodes in different layers and nodes in each layer may be needed. For example, the paper \cite{stavrinides2019hybrid} presented a heuristic allocation and scheduling approach to minimize the execution time of independent tasks of IoT applications in the form of workflows. It is considered multiple fog and cloud servers in the system model while they are fully connected by a virtual network. A latency-aware application module placement of workflow-based applications in a fog computing environment with multiple fog servers and one cloud server is presented in \cite{mahmud2018latency} that meets the diverse service delivery latency and amount of data signals to be processed in per unit of time for different applications. A heuristic approach for allocating multiple workflows in a fog computing system with multiple edge/fog and cloud servers is presented in \cite{goudarzi2020application}. A meta-heuristic approach is introduced in \cite{xu2019computation} to address the offloading problem for multiple workflows and is studied to minimize time and energy consumption. The paper \cite{de2019resource} presents a heuristic approach to minimize the time and monetary cost of executing multiple workflows in a fog computing system. This paper also considered the fog and cloud servers in the form of a connected graph. The paper \cite{ijaz2021energy}, presents a heuristic approach to schedule delay-sensitive IoT applications in the form of workflows in a fog computing environment. An offloading and scheduling solution for workflows in fog computing is presented in \cite{de2020multi}.\\
On the other hand, large-scale scientific applications may comprise a set of interrelated workflows called a workflow ensemble \cite{bharathi2008characterization, genez2017robust, jiang2015executing, malawski2015algorithms, taylor2007workflows, genez2016flexible}. Workflows in an ensemble typically have a similar structure, and the differences are in their sizes (number of tasks), input data, and individual computational task sizes \cite{bharathi2008characterization, taylor2007workflows}. There are researchers on managing workflow ensembles in cloud computing \cite{malawski2015algorithms,genez2017robust,genez2016flexible, pietri2013energy}, also \cite{siar2021offloading} studied offloading and allocation of workflows in a fog computing environment. These approaches are considering multiple fog and cloud servers in their environment. So, assigning the tasks to different servers requires sending dependency data among them. In this case, devices in different layers must communicate with each other, which is practical in a graph topology. Therefore, a simulation tool that is unable to model the graph topologies cannot simulate and evaluate most of the presented studies.

\subsection{Simulation tools}\label{simtool}
CloudSim, \cite{calheiros2011cloudsim} is a standard network simulator for simulating scheduling and resource management algorithms of applications in the form of independent tasks. This simulator provides numerous entities that are necessary to simulate a cloud environment. It is an open-source toolkit where different aspects can be reprogrammed. Despite the tooling provided by CloudSim and its flexibility to be partially reprogrammed, it cannot be used to simulate the execution of a workflow in a fog computing environment. This simulator supports modeling independent tasks and their properties, such as the number of instructions to be completed. However, it lacks support for properties such as the dependency between a group of tasks which is crucial for modeling a workflow. Another limitation is the severity of the arrival time of each task, such that all tasks must enter the simulation environment at time zero.
iFogSim simulator \cite{gupta2017ifogsim} is developed upon CloudSim to add the necessary features for simulating a fog computing environment to that one. In contrast to CloudSim, iFogSim has richer capabilities for modeling the dependent tasks and their dependencies. Its computational model is essentially a graph in which each vertex is a task, and each edge models the data flow from a source task to a destination task. Since a workflow (i.e., DAG) is a tree of tasks, we can consider all workflows as a subset of all computational models supported by the iFogSim. Additionally, iFogSim supports various devices, including sensors, actuators, edge, fog, and cloud devices. These features make iFogSim an excellent choice for modeling the execution of applications in a fog computing environment. However, the importance of these features is diminished by the strict modeling of network topology. In iFogSim, network devices can only form a tree-like topology. Also, since there are no routing tables, each device uses broadcast to send its message to the destination, making it unsuitable for modeling all real-world networks. Its computational model lacks support for grouping the tasks into distinct workflows. Also, there is no support for defining parameters like the deadline for each task or the application. iFogSim2, \cite{mahmud2022ifogsim2} an enhanced version of iFogSim, aimed to address certain issues in iFogSim. It achieved this by introducing a set of simulation models for mobility-aware application migration, dynamic distributed cluster formation, and microservice orchestration. These contributions positively influenced the quality of service and resource utilization within the simulation. While these enhancements were impactful, the underlying computational and network models remained largely unchanged. As a result, iFogSim2 inherits the primary limitations of its predecessor.
WorkflowSim, \cite{chen2012workflowsim} is also developed upon CloudSim but with a different goal: to simulate the execution of scientific workflows in distributed environments. This simulator enables the user to model scientific workflows providing a directed acyclic graph as its computational model. WorkflowSim takes different delays such as queuing delay, postscript delay, and workflow engine delay into account during the simulation. It can model random failures in the system and respond to them according to users’ configuration, making it an excellent choice for modeling fault-tolerant systems and algorithms.
One of the shortcomings of this simulator is the lack of support for concurrent execution of multiple workflows. All tasks in each workflow will mix after applying to the workflow engine component, and it cannot be distinguished each task corresponds to which workflow. Also, WorkflowSim supports a star-like topology for networking in which the central node is the broker and other nodes connected to it are data-centers. These limitations make WorkflowSim unsuitable for decentralized schemes.
FogWorkflowSim, \cite{liu2019fogworkflowsim} is developed on top of both iFogSim and WorkflowSim. This simulator tries to use the network model of iFogSim and integrate it with the computation model of WorkflowSim. Therefore, it is subjected to the shortcomings of the network model of iFogSim and the computation model of WorkflowSim. 
IoTSim-Edge, \cite{jha2020iotsim} has also developed upon CloudSim to accurately model protocols, devices, and scenarios that take part in executing an application on the edge of the network. In this simulator, numerous features are available to model different aspects of the edge network, such as modeling the movement of devices at the edge of the network and their dynamic connection to each cell tower. But, just like iFogSim, this simulator only supports a tree-like topology for networking and suffers from the same constraints as that one. The limitations of the computational model of WorkflowSim also exist here. EdgeCloudSim simulator, \cite{sonmez2017EdgeCloudSim} also extends CloudSim with a modular architecture to assess the performance of edge computing systems. Notably, it excels in specialized network modeling for WLAN and WAN scenarios, integrates a device mobility model, and features a realistic, tunable load generator and network delay model. Compared to IoTSim-Edge, which focuses on precise modeling of edge protocols and devices, EdgeCloudSim stands out with its modular design and dedicated network modeling capabilities. However, as it relies on CloudSim capabilities for its computational model, EdgeCloudSim faces the same limitations that exist in the CloudSim computational model and cannot model the scientific workflows that are heavily used in IoT applications.\\
After investigating standard simulators in the context of cloud, fog, and edge simulation, we concluded that CloudSim is the best choice as the base simulation framework to develop a simulation toolkit for modeling resource management problems. Our goal is to reach a simulator where the computational model can support any type of workflow-based application. Also, the network model can support both centralized and decentralized resource management algorithms by considering most of the topologies introduced for the distributed environments of cloud, edge, and fog computing.

\section{Architecture}
\label{arch}
Each discrete-event simulation framework requires tools to model entities and events. Also, there needs to be an interface for sending and receiving events. The simulation framework manages each entity's life cycle and the generation and delivery of events.
When the necessary infrastructure is provided, one can build higher-level tools on top of it. For example, many components such as data-centers, brokers, virtual machines, and processing cores need to be modeled for a fog computing environment. Also, other concepts like tasks, files, and network packets may not be an entity but need to be modeled.
In addition, other concepts don't take a physical form but are present in a resource management and scheduling simulation environment, such as provisioning algorithms and schedulers.
It is efficient to rely on a well-established discrete event simulation framework, and develop our desired simulator on top of it. It can provide us with the basic tooling that we need, and we can focus on implementing higher-level features.\\ 
iFogSim’s network model simulates tree topologies, and its computational model considers workflows. However, the computational model in the iFogSim is based on the distributed data flow (DDF) model, and converting this model to a workflow as a directed acyclic graph (DAG) format is more complex than extending the computational model of CloudSim to support workflow-based applications.  We picked CloudSim as the base simulator for WIDESim. The main reason for this decision is the negligible amount of friction that CloudSim introduces for the extension. Also, according to section \ref{rel}, all other state-of-the-art simulators are based on CloudSim, indicating the flexibility of this simulation toolkit to be customized for different purposes. CloudSim provides us with all the necessary tooling for discrete-event simulation in cloud computing. Its computation model is rudimentary but quite easy to extend. Its network model contains all the essential features needed to operate in a network. So the challenge lies in making every device in the network aware of the topology and supporting workflow-based applications. In the following section, we discuss the relationship between the architecture of CloudSim and that of WIDESim.

\subsection{The proposed simulator}
\begin{figure}[h]
\centering
\includegraphics[width=0.6\textwidth]{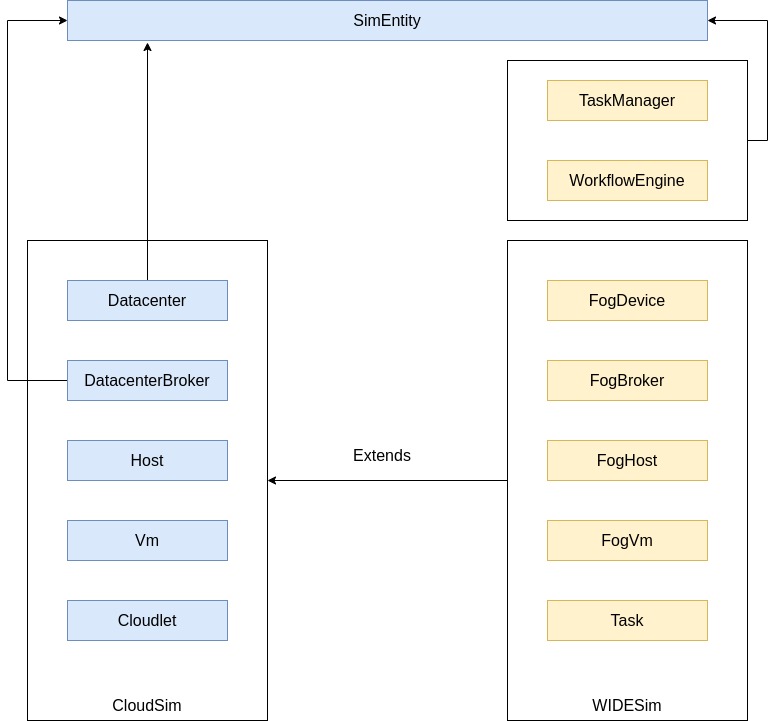}
\caption{The relationship between classes in WIDESim and CloudSim}
\label{fig:widesim-cloudsim-rel}
\end{figure}

We attempted to implement WIDESim to take inspiration from the discussed simulators in section \ref{rel} and inherit their merits while improving their shortcomings. Figure \ref{fig:widesim-cloudsim-rel} represents the relationship between the architecture of WIDESim and CloudSim in terms of their classes. The architecture of WIDESim consists of two primary categories: the network model and the computational model. Components such as FogDevice, FogHost, and FogVM are part of the network model and are responsible for managing data movement between devices and the actual processing of tasks. Conversely, components like TaskManager, WorkflowEngine, and FogBroker belong to the computational model, overseeing task relations and making decisions on releasing and assigning tasks to devices based on user policies. The details of each component class are presented in the following.

\subsubsection{FogDevice}
\begin{figure}[h]
\centering
\includegraphics[width=0.6\textwidth]{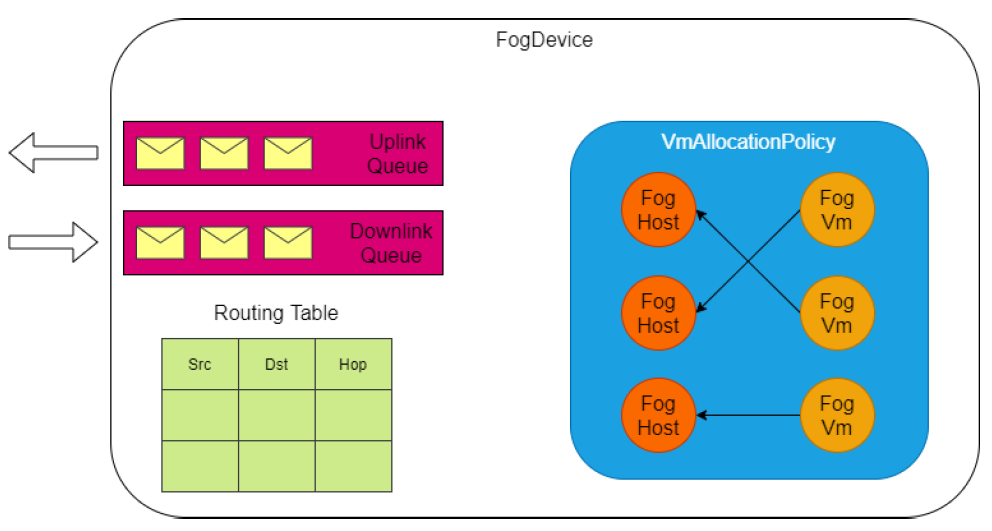}
\caption{An overview of FogDevice class}
\label{fig:fogDevice}
\end{figure}

The FogDevice class models a device in a fog environment network. Therefore, it inherits the properties of a data-center from the Datacenter class in CloudSim, which is also an inheritance of the SimEntity class. So, the FogDevice class is an entity that can receive, process, and send simulation events. This class is modeling a data-center having knowledge from the network topology. This knowledge helps fog devices to communicate with each other. The FogDevice class has three major components of links, routing table, and VM allocation policy. Figure \ref{fig:fogDevice} represents the main components of this class. The routing table consists of triples in the form of (source id, destination id, next-hop), which indicates which fog device should be the next hop in the route to any destination. Using routing table, the fog devices can route the received packages in the network. When a fog device receives a message, check if it is the destination of this message, or not. If yes, it will open the message, otherwise, the fog device will send the message to a neighbor node, according to its routing table. This process repeats until the destination device receives the message. Also, there is a download and upload link in each fog device. Each link can receive each packet of data sequentially, and other packets will be stored in an FCFS queue. And finally, the VM allocation policy maps the VMs created in the FogDevice to a FogHost in this FogDevice. Users can customize the VM allocation policies for this class.

\subsubsection{FogHost}
This class is a subclass of the Host in Cloudsim without any unique additions. It was created to expose some private properties of the Host class that are needed for the dynamic nature of the WIDESim simulator.

\subsubsection{FogVM}
This class is a subclass of the Vm in Cloudsim without any special additions. It was created to expose some private properties of the Vm class that are needed to support dynamic resource management in the WIDESim.

\subsubsection{Task}
\begin{figure}[h]
\centering
\includegraphics[width=0.6
\textwidth]{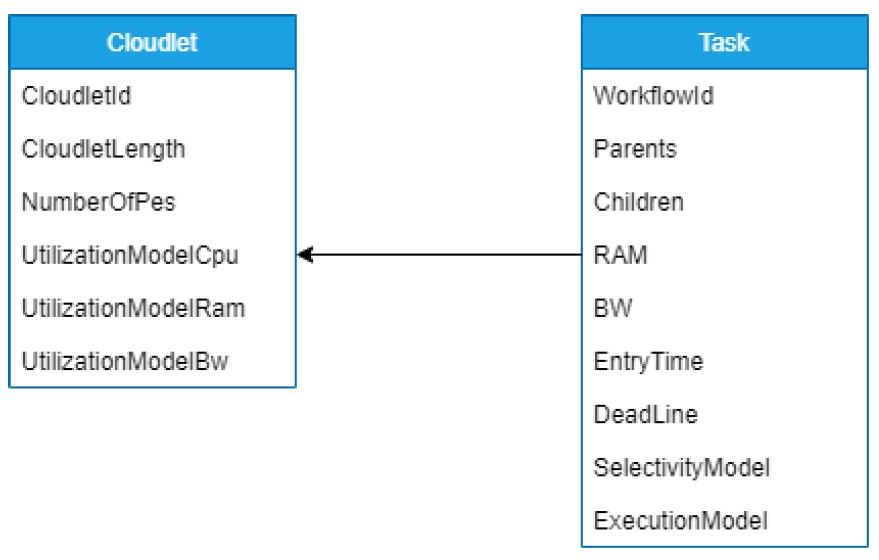}
\caption{The relationship between Task and Cloudlet classes}
\label{fig:task}
\end{figure}

Task represents a job in the WIDESim, while it is aware of its position in a workflow. This class is inherited from the Cloudlet class in Cloudsim, which models workload. However, the Cloudlet class has limitations for modeling a task in a workflow according to the WIDESim purpose. So, we inserted some features to the Cloudlet linking each task in a workflow to its parent(s) and children. Each task also carries the id of the workflow belongs to. Other attributes specify how the task will behave when its execution is completed. For example, the selectivity model dictates how the task will produce its output based on its received inputs. The execution model specifies how the task will continue its life cycle after its execution is completed. For example, the task can behave periodically and restart its execution after a fixed period. Each task also carries other specific attributes like its entry time and deadline. Figure \ref{fig:task} represents Cloudlet's features and the features inserted into it in the Task class of WIDESim.

\subsubsection{TaskManager}
TaskManager is a subclass of SimEntity in the CloudSim. So, it can receive, process, and send events. TaskManager is responsible for releasing tasks into the simulation environment based on their entry time. TaskManager itself doesn't know the dependencies between tasks and only looks at their entry time. This entity has access to all input tasks and sorts them in a queue according to their entry time, and releases a task when the simulation clock is equivalent to this task's entry time. The released tasks are sent to the WorkflowEngine which knows the dependency between tasks. TaskManager is not a physical component in the network and only simulates tasks' dynamic arrival into the simulation environment.

\subsubsection{WorkflowEngine}
WorkflowEngine is inherited from SimEntity class in CloudSim. This entity resolves dependencies among tasks and has absolute knowledge of the structure of all workflows submitted to the simulation environment. It receives incoming tasks from the TaskManager. Each received task enters into a queue until the execution of all of its parents is completed. Then, this task is ready for scheduling. So, WorkflowEngine will release this task and pass it to the FogBroker for scheduling.

\subsubsection{FogBroker}
This class may be the centerpiece of WIDESim. It is also an inheritance of SimEntity class in CloudSim. FogBroker knows the network's topology, mapping tasks to VMs, and mapping VMs to FogDevices. Also, it keeps track of the life cycle of each task and generates the necessary messages and events based on each task's execution and selectivity model. This class also keeps track of the life cycle of different VMs and can create and destroy VMs based on a provisioning algorithm. In this case, three task and VM management methods are considered in this class as follows:\\
\begin{itemize}
    \item {VmToFogDeviceMapper: This method determines the allocation of VMs to fog devices.}
    \item{TaskToVmMapper: The manner of mapping tasks to VMs controls by this method.}
    \item{VMProvisioner: This method can change VM’s status by deciding which VMs are provisioned, failed, or destroyed.}
\end{itemize}
All the above methods are rewritable and can be customized according to the user’s purpose. 

\subsubsection{Input Files}
\begin{figure}[h]
\centering
\includegraphics[width=0.3\textwidth]{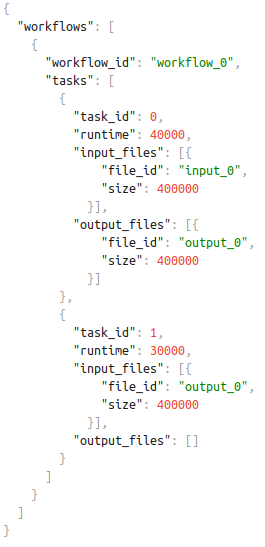}
\caption{An example of Workflows.json}
\label{fig:workflows-json-example}
\end{figure}

In addition to the introduced classes, three input files are considered for describing different aspects of the simulation environment including the input application, network model and user-specified configurations of the simulation environment. These files which are explained in the following are parsed by their designated parser:

\textbf{Workflows.json}: This file contains descriptions of input workflows that are supposed to be run in the simulator. At the root level, there is the "workflows" tag, which is an array of workflows. Any number of workflows can be defined in this file. Each workflow contains a "workflow\_id" and an array of tasks. The definition of tasks of each workflow begins with the "tasks" tag. Each task has an id, a runtime that indicates the number of instructions needed to complete this task, an array of input files, and an array of output files. This file parses by WorkflowParse in WIDESim. An example of this file is depicted in figure \ref{fig:workflows-json-example}. This example models a single workflow with id 0, composed of two tasks with ids 0 and 1, and runtimes of 40000 and 30000 million instructions (MIP), respectively. The input file of task with id 1 is dependent on the output of task with id 0.

\begin{figure}[h]
\centering
\includegraphics[width=0.3\textwidth]{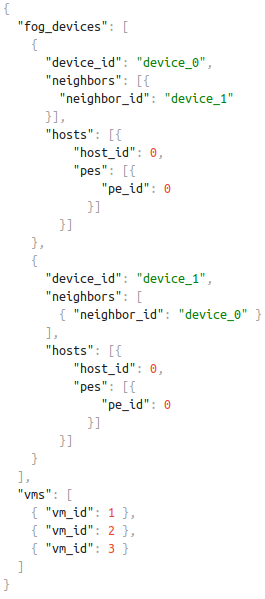}
\caption{An example of Topology.json}
\label{fig:toplogy-json-example}
\end{figure}

\textbf{Topology.json}: This file is parsed by the TopologyParser in WIDESim, and describes the network architecture and its topology. Using this file, the user can define the specifications of fog devices available in the network and the connections between them. All properties of a network environment such as fog devices, hosts, processing cores, and virtual machines are configurable through this file. Any network topology can be defined using this file. As shown in figure \ref{fig:toplogy-json-example} at the root level of this file is the "fog\_devices" tag, which is an array of FogDevices class. Each FogDevice has an id and a list of neighbors to which this device is connected. Neighbors are instance of FogDevice too. Also, it has a list of hosts belonging to this device. Each host is an instance of FogHost class and has an id and an array of processing cores. Another root level tag is the "vms" tag which lists the virtual machines created in the simulation environment. It represents FogVM class. The complete list of tags can be found in the \href{https://github.com/ARH80/WIDESim/blob/master/src/main/java/widesim/parse/topology/Parser.java}{\texttt{src/main/java/widesim/parse/topology/Parser.java}} file, within the \texttt{Tags} class. Figure \ref{fig:toplogy-json-example} represents an example of Topology.json when there are two fog devices with IDs 0 and 1 in the system, that are connected directly. There is one host with one processing core in each fog device. Also, three VMs with IDs 0 through 3 are defined in this file. 
More examples of this file can be found in the \href{https://github.com/ARH80/WIDESim/blob/master/src/main/resources/topologies}{\texttt{src/main/resources/topologies/}} directory.

\begin{figure}[h]
\centering
\includegraphics[width=0.5\textwidth]{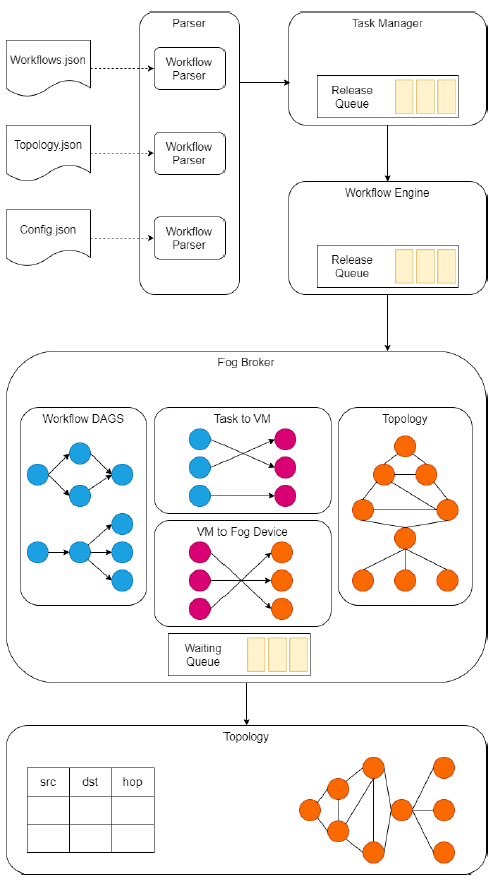}
\caption{An overview of the relationship between WIDESim's entities and their functionality}
\label{fig:arch}
\end{figure}

\textbf{Config.json}: This file is parsed by the ConfigParser in WIDESim and provides the user-specified configurations about the behavior of the simulation environment.  For example, we can define how the entities must react when the deadline of a task is reached while it is not completed.\\
More examples for these files can be found from \href{https://github.com/ARH80/WIDESim}{\texttt{our Github repository}}. A separate parser is used to parse and analyze each of these files and generate the necessary information for the simulation environment. This information is distributed between TaskManager, FogBroker, and FogDevices once needed. The tasks extracted by WorkflowParser are passed to the TaskManager. Then, this component passes the tasks to the WorkflowEngine according to their entrance time. Finally,
the tasks are sent to the FogBroker according to their data dependencies in their associated workflow. FogBroker routes
the tasks to a FogDevice for scheduling, knowing the network’s topology. An overview of the relationship between WIDESim's entities and their functionality is presented in figure \ref{fig:arch}.

\subsubsection{Failure Generator}

Failure generator can be used to inject task failure at each simulation run. 
Then the failed tasks will be rescheduled by the Broker.
Failure generator randomly selects a task and fails it based on the specfied random distribution 
and the corresponding parameters.
Currently, WIDESim supports Normal, Log-Normal, Gamma, and Weibull distributions. By enabling the failure, the failure generator will be called after the execution of each task; 
then a column will be added to the output table to show the failure time(s) of each task.

Failure generator examples and usage can be found in \href{https://github.com/ARH80/WIDESim/tree/master/src/main/java/widesim/examples/fail}{\texttt{src/main/java/widesim/examples/fail}} directory. 
The output of the examples are available in \href{https://github.com/ARH80/WIDESim/tree/master/outputs/fail}{\texttt{outputs/fail}} directory.

\subsection{Summary of characteristics}
According to the architecture of WIDESim, we can summarize its features compared to the reviewed simulators in table \ref{table:comparison}. As shown in this table, network simulators discussed in section \ref{rel} do not provide the necessary features needed to simulate the concurrent execution of workflows in a fog computing environment, while WIDESim provides this
feature. WIDESim enables separating workflows during execution, allowing defining different QoS parameters for each
workflow. Using this feature, our developed simulator can model any application in the form of scientific workflows.
None of the reviewed simulators support the dynamic entrance of jobs to the system, while WIDESim supports this feature
to enable dynamic resource management and scheduling. Another distinctive feature of the developed simulator is
the ability to support any network topology. WIDESim uses the routing tables on each fog device to send a message
directly between neighboring nodes, and there are no limitations for communications between nodes. On the other hand,
the topology supported in the reviewed simulators is the star or tree. This feature of WIDESim enables the modeling of
broader architectures of edge paradigms and IoT use cases.

\begin{table}[]

\begin{adjustbox}{width=1.\textwidth}
\begin{tabular}{|c|c|c|c|c|c|c|}
\hline
 & CloudSim  & iFogSim   & \makecell{Workflow\\Sim}  & \makecell{Fog\\WorkflowSim}  & \makecell{IoTSim\\Edge}  & WIDESim    \\
\hline
Working with workflows                              & x    & \checkmark & \checkmark & \checkmark & \checkmark & \checkmark  \\
\hline
Concurrent execution  & \checkmark & \checkmark & \checkmark & \checkmark & \checkmark & \checkmark  \\
\hline
Grouping   & x & x    & x  & x   & x  & \checkmark  \\
\hline
\makecell{Define specific parameters\\ for each workflow}  & x    & x & x    & x  & x   & \checkmark  \\
\hline
Dynamic arrival of tasks   & x  & x & x & x  & x & \checkmark  \\
\hline
\makecell{Dynamic scheduling \\and provisioning}  & x  & x  & x  & x   & x & \checkmark  \\
\hline
Network Topology  & Star & Tree  & Star      & Tree  & Star            & Graph \\
\hline
\end{tabular}
\end{adjustbox}

\caption{Specification of the network layer used in the evaluation\label{table:comparison}}
\end{table}

\section{Evaluation and Validation}
\label{eval}

To study the validity of WIDESim's performance, we conducted a two-part evaluation under different conditions. In the first part, we compared it with standard simulators. In these evaluations, the simulation time and energy consumption of the proposed simulator are compared to standard simulators, which their accuracy proved. If the results of two simulators are close to each other in an identical scenario, the functionalities are also the same. The contribution of the WIDESim is in both the application and network model. So, for exact comparison, the validation of the WIDESim in this part has been divided into two sections. At first, we will keep the network model simple to evaluate the proposed simulator against WorkflowSim, in terms of the computation model. Then we add more nodes to the network to verify the validity of the proposed simulator’s network model by comparing it with iFogSim. The network model of IoTSim-Edge is more restricted than iFogSim, and its computation model is on par with WorkflowSim. Also, FogWorklowSim has the network model of iFogSim and the computation model of WorkflowSim. So comparing WIDESim with IoTSim-Edge and FogWorkflowSim does not yield a new scenario that could not be validated during our comparison with iFogSim and WorkflowSim.

In the second part of our evaluations, our focus is solely on showcasing WIDESim's capabilities in both network and computation models without imposing restrictions on various simulation factors. We primarily measure completion time and overall energy consumption as key performance metrics for this part. To achieve a comprehensive evaluation, we introduce three different network topologies that are more complicated than those used in the first part. These complex topologies effectively demonstrate WIDESim's support for a wide range of network structures and its suitability for various real-life IoT scenarios. Furthermore, we thoroughly assess WIDESim's robust computation model through three distinct sections. In the first section, we evaluate WIDESim's performance with single scientific workflows, including Cybershake, Montage, Sipht, Inspiral, and Epigenomics, as input. In the second section, we conduct experiments using multiple workflow inputs, which consist of batches of different workflows of varying sizes. Finally, in the third section, we evaluate WIDESim with workflow ensembles, which are collections of workflows of the same type but with different sizes. These sections collectively confirm WIDESim's support for processing diverse workflow-based applicationand edge-based network architectures

To maintain our focus on evaluating the fundamental capabilities of the network and computational models, we intentionally disabled error generation and fault tolerance features in all experiments. It's also important to note that energy consumption in all the simulations is calculated as the product of the average power of the devices present in the network throughout the entire simulation duration. The measured time is in seconds, and energy (power) consumption is in joules (watts).

\subsection{Comparison with presented simulators}
In this section, we are going to compare WIDESim with some of the most common simulation tools available.

\subsubsection{Computational model evaluation}
\begin{figure}[h]
\centering
\includegraphics[width=0.6\textwidth]{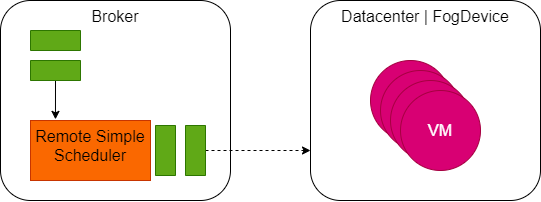}
\caption{The network topology used in comparing WIDESim with WorkflowSim}
\label{fig:compute_net_model}
\end{figure}

\begin{figure}[h]
\centering
\includegraphics[width=0.5\textwidth]{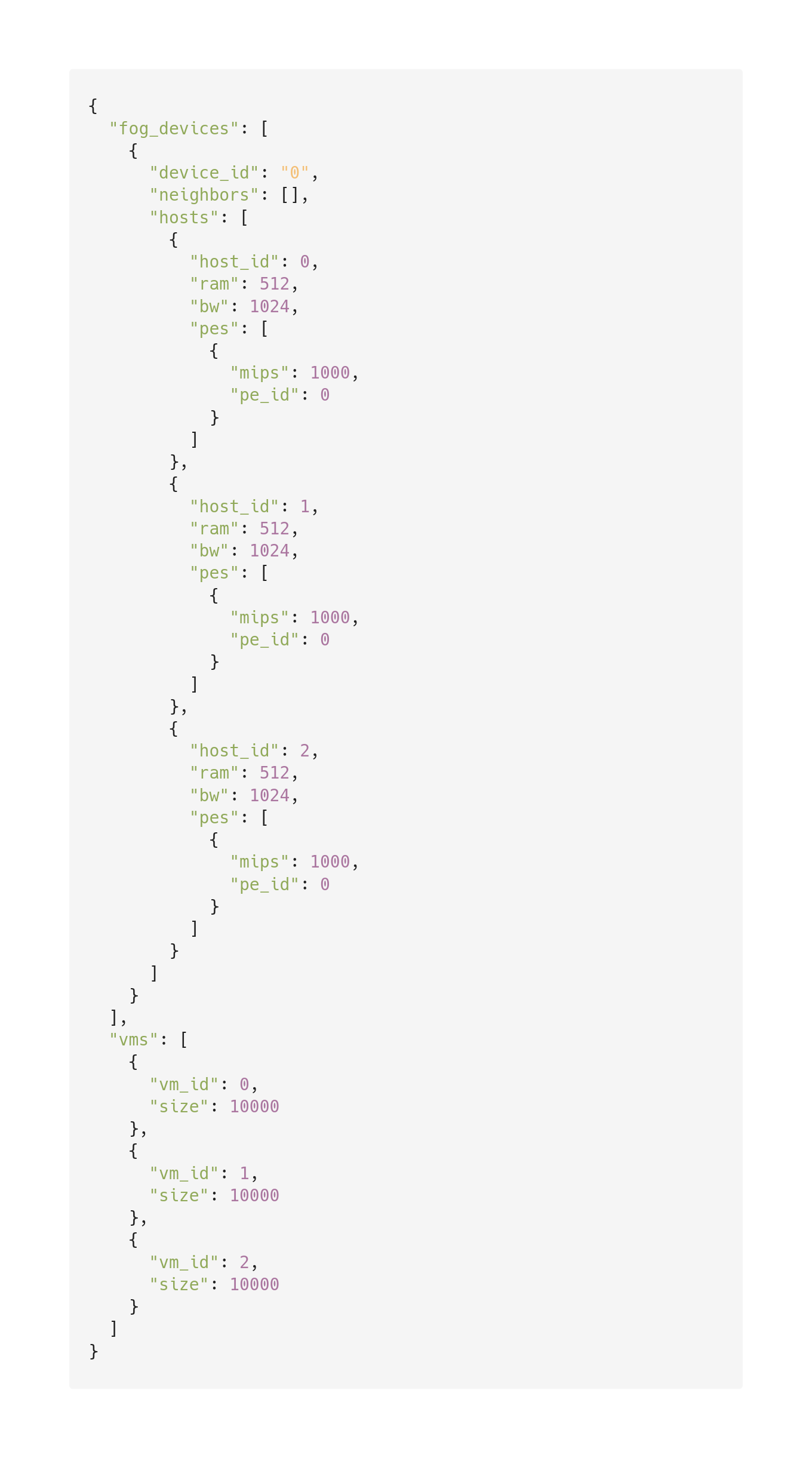}
\caption{The JSON file of network topology used in comparing WIDESim with WorkflowSim}
\label{fig:eval-topology}
\end{figure}

To evaluate the computational model of the proposed simulator, WorkflowSim has been chosen as a reference. This simulator has one of the richest computational models among existing simulators, in terms of supporting workflow-based applications. This simulator is supporting execution of workflows, similar to our simulator. This issue caused a close similarity between the computational model of WorkflowSim and the proposed simulator. However, the similarities end here, and we have to limit the network model of our simulator immensely to make it compatible with WorkflowSim. This is due to WorkflowSim's restriction to a star network topology, wherein a central broker resides at the core, and the data-centers are directly linked solely to this central node. Therefore, the network depicted in figure \ref{fig:compute_net_model} has been chosen as the underlying topology.  The exact specification of the simulation environment in this comparison is shown in table \ref{table:compute_net_spec}. The network topology includes a single data-center, and the resources of this data-center will be divided equally between three VMs.\\
The JSON file for this topology is depicted in figure \ref{fig:eval-topology}.

To ensure the validity and fairness of the evaluations, we have configured all parameters identically across both simulators. Furthermore, any extra features inherent to each simulator, such as clustering or fault tolerance, have been deliberately deactivated. Both simulators incorporate a customizable mapping mechanism: the TaskToVmMapper class in WIDESim and the BaseSchedulingAlgorithm class in WorkflowSim that can be explicitly specified within WorkflowSim's Parameters class. These mechanisms are responsible for assigning tasks to specific VMs before their submission to the broker. Moreover, these mechanisms offer high reconfigurability, allowing users to implement and experiment with their own mapping algorithms across both simulators. In this paper, we have employed two distinct mapping algorithms to enhance the precision and comprehensiveness of our computational model evaluation. The first algorithm is a straightforward round-robin (RR) mapper. This approach involves uniformly distributing all ready tasks among all the free VMs, leading to a nearly equal allocation of tasks to each VM. Additionally, each VM executes its assigned tasks using a time-shared methodology. A second algorithm, an FCFS mapper, is also utilized. This algorithm prioritizes assigning ready tasks with earlier entry times to available VMs first. It then waits for the completion of a task and the availability of the associated VM before proceeding to assign the remaining ready tasks. Our evaluation encompasses both WIDESim and WorkflowSim, including both of these mapping algorithms to comprehensively assess their performance.\\

\begin{figure}
\centering
\begin{minipage}{0.31\textwidth}
\includegraphics[width=\linewidth]{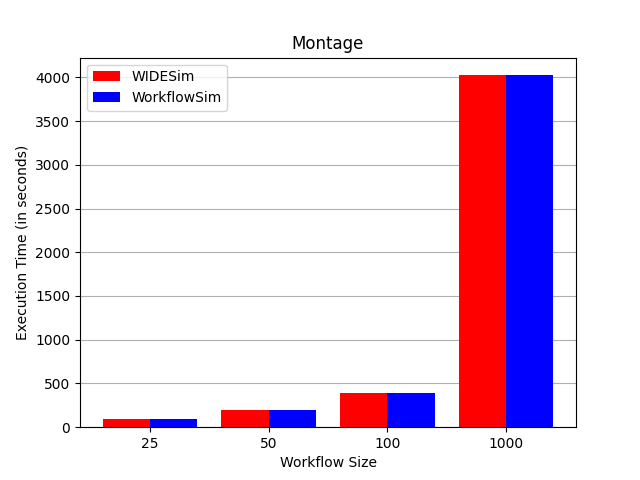}
\subcaption{Montage}
         \label{fig:montage_RR_result}
\end{minipage}
\hspace{1mm} 
\begin{minipage}{0.31\textwidth}
\includegraphics[width=\linewidth]{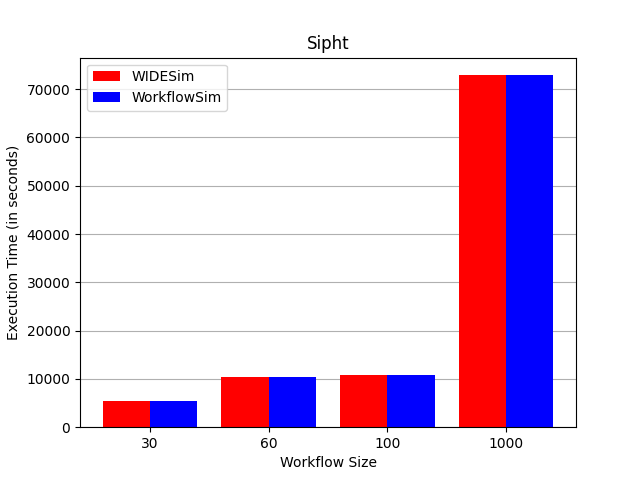}
\subcaption{Sipht}
         \label{fig:sipht_RR_result}
\end{minipage}

\medskip
\begin{minipage}{0.31\textwidth}
\includegraphics[width=\linewidth]{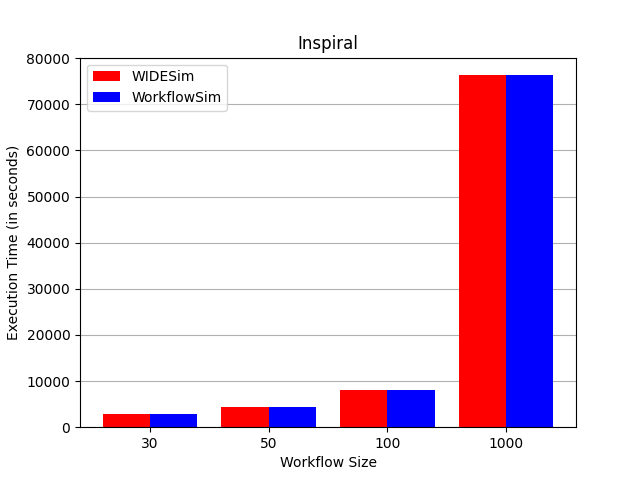}
\subcaption{Inspiral}
         \label{fig:inspiral_RR_result}
\end{minipage}
\hspace*{\fill}
\begin{minipage}{0.31\textwidth}
\includegraphics[width=\linewidth]{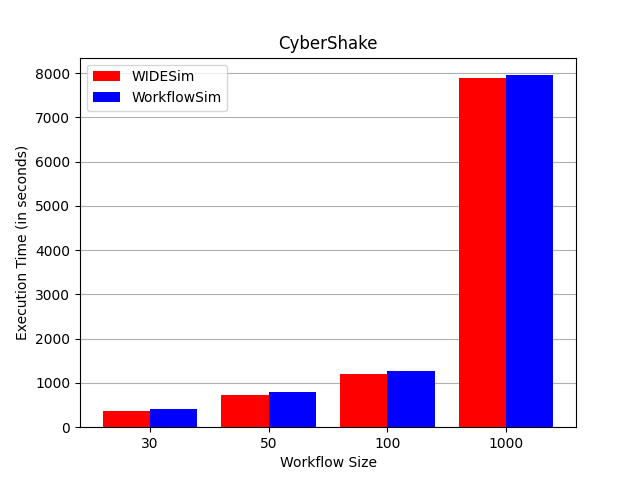}
\subcaption{Cybershake} \label{fig:cybershake_RR_result}
\end{minipage}
\hspace*{\fill}
\begin{minipage}{0.31\textwidth}
\includegraphics[width=\linewidth]{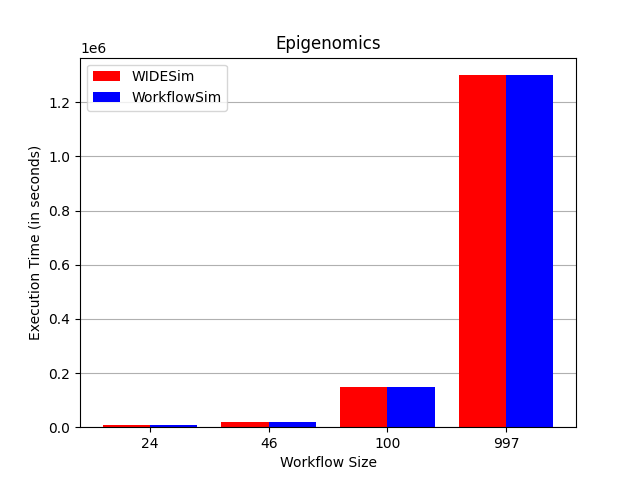}
\subcaption{Epigenomics} \label{fig:epigenomics_RR_result}
\end{minipage}

\caption{Results of evaluating the computational model of WIDESim compared with WorkflowSim (round-robin mapper)}
\label{fig:workflow_RR_results}
\end{figure}

\begin{figure}
\centering
\begin{minipage}{0.31\textwidth}
\includegraphics[width=\linewidth]{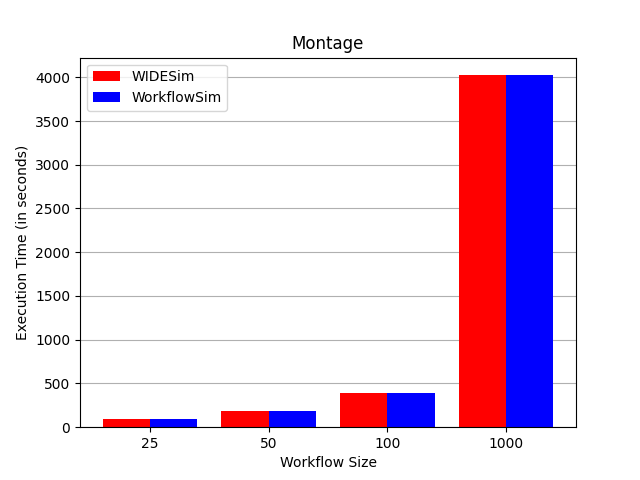}
\subcaption{Montage}
         \label{fig:montage_FCFS_result}
\end{minipage}
\hspace{1mm} 
\begin{minipage}{0.31\textwidth}
\includegraphics[width=\linewidth]{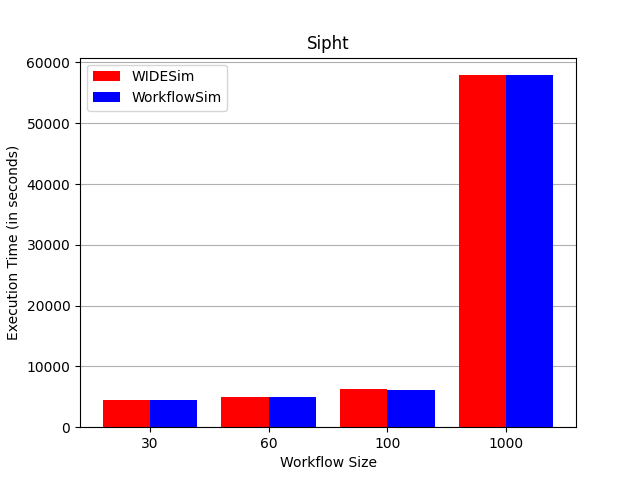}
\subcaption{Sipht}
         \label{fig:sipht_FCFS_result}
\end{minipage}

\medskip
\begin{minipage}{0.31\textwidth}
\includegraphics[width=\linewidth]{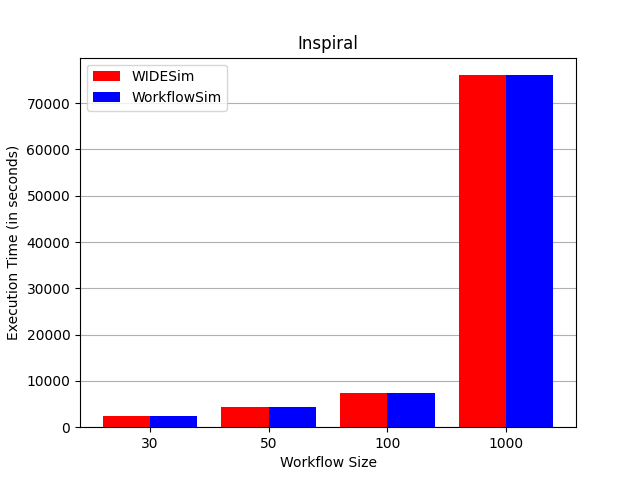}
\subcaption{Inspiral}
         \label{fig:inspiral_FCFS_result}
\end{minipage}
\hspace*{\fill}
\begin{minipage}{0.31\textwidth}
\includegraphics[width=\linewidth]{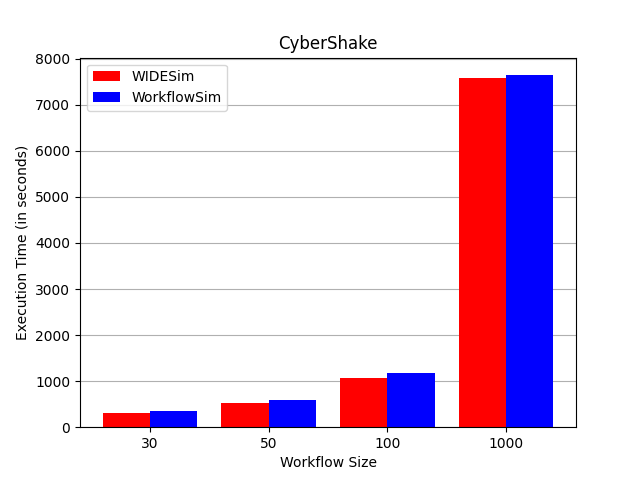}
\subcaption{Cybershake} \label{fig:cybershake_FCFS_result}
\end{minipage}
\hspace*{\fill}
\begin{minipage}{0.31\textwidth}
\includegraphics[width=\linewidth]{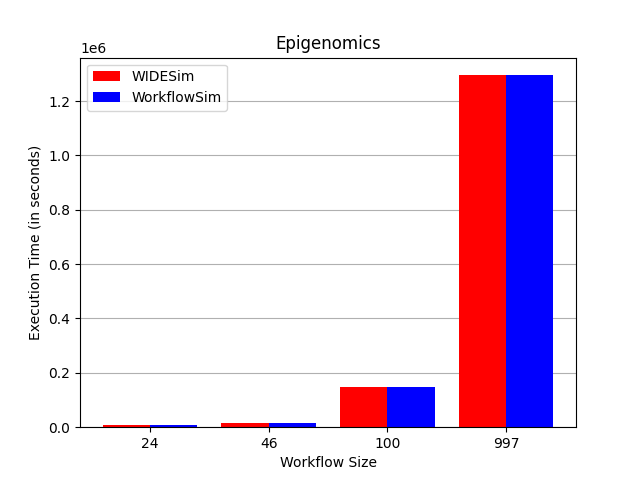}
\subcaption{Epigenomics} \label{fig:epigenomics_FCFS_result}
\end{minipage}

\caption{Results of evaluating the computational model of WIDESim compared with WorkflowSim (FCFS mapper)}
\label{fig:workflow_FCFS_results}
\end{figure}

The execution time of WorkflowSim and WIDESim for five scientific workflows of Montage, Sipht, Inspiral, Cybershake, and Epigenomic \cite{juve2013characterizing} and both mapping algorithms, RR and FCFS,  are presented in table \ref{table:compute_results}. Visual comparisons between WIDESim and WorkflowSim utilizing the round-robin and FCFS mapping algorithms are illustrated in figures \ref{fig:workflow_RR_results} and \ref{fig:workflow_FCFS_results} respectively. The results of both simulators are either equal or incredibly close to each other. These negligible differences arise from inherent structural differences between the two simulators that cannot be manipulated in an experiment. For example, WorkflowSim is a cloud computing simulator, and file transfer times between hosts within the same data-center are not instantaneous. Conversely, WIDESim functions as a fog computing simulator, so there are no file transfer times between hosts within the same fog device. Upon intentionally setting the file transfer time to 0 in WorkflowSim, the resulting outcomes aligned with those of WIDESim across all analyzed workflows. All the execution results are available in the outputs/computational\_model\_evaluation directory in the Github repository. These results prove that despite inserted features in our simulator, its accuracy is similar to WorkflowSim. Since the considered scientific workflows in this comparison have a different structure in terms of the graph and the tasks' specifications, the results further show that our simulator can maintain a similar accuracy compared to WorkflowSim, for different types of scientific workflows. Therefore, our simulator can be used in any scenario where WorkflowSim falls short, as discussed in section \ref{arch}, such as all forms of scientific workflow-based applications and decentralized resource management and scheduling. 

\begin{table}
\centering
\begin{tabular}{|c|c|} 
 \hline
 \textbf{Parameter} & \textbf{Value} \\ 
 \hline
 Number of VMs & 3 \\ 
 \hline
 VM size & 10000 MB \\
 \hline
 RAM & 512 MB \\ 
 \hline
 Bandwidth & 1024 MB/s \\
 \hline
 MIPS & 1000 MIPS \\ 
 \hline
 Number of processing cores & 1 \\
 \hline
 Scheduling algorithm & CloudletSchedulerTimeShared \\ 
 \hline
 Number of data-centers (fog devices) & 1 \\
 \hline
 Number of hosts & 3 \\ 
 \hline
 Number of processing cores for each host & 1 \\
 \hline
 MIPS of each host & 1000 MIPS \\ 
 \hline
 Processing core Provisioning algorithm & PeProvisionerSimple \\
 \hline
 RAM of each host & 512 MB \\ 
 \hline
 RAM Provisioning algorithm & RamProvisionerSimple \\
 \hline
 Bandwidth of each host & 1024 MB/s \\
 \hline
 Bandwidth Provisioning algorithm & BwProvisionerSimple \\ 
 \hline
 VM allocation policy & VmAllocationPolicySimple \\
 \hline
 VM scheduler algorithm & VmSchedulerTimeShared \\
 \hline
\end{tabular}
\caption{Specification of the simulation environment in evaluating WIDESim vs WorkflowSim\label{table:compute_net_spec}}
\end{table}

\begin{table}
\centering
\begin{adjustbox}{width=1\textwidth} \begin{tabular}{|c|c|c|c|c|} 
 \hline
 \textbf{Workflow} & \textbf{WIDESim (RR)} & \textbf{WorkflowSim (RR)} & \textbf{WIDESim (FCFS)} & \textbf{WorkflowSim (FCFS)} \\ 
 \hline
 Cybershake 30 & 360.99 & 400.54 & 317.97 & 357.8 \\
 \hline
 Cybershake 50 & 719.57 & 791.25 & 522.78 & 580.32 \\
 \hline
 Cybershake 100 & 1200.16 & 1272.16 & 1075.69 & 1181.61 \\
 \hline
 Cybershake 1000 & 7893.48 & 7952.86 & 7588.47 & 7642.43 \\
 \hline
 Epigenomics 24 & 8793.33 & 8796.44 & 8412.78 & 8416 \\
 \hline
 Epigenomics 46 & 17740.24 & 17742.25 & 15395.89 & 15399 \\
 \hline
 Epigenomics 100 & 147493.22 & 147499.43 & 146196.31 & 146183.78 \\
 \hline
 Epigenomics 997 & 1300291.68 & 1300298.37 & 1295050.89 & 1294140.58 \\
 \hline
 Inspiral 30 & 2880.56 & 2880.78 & 2429.92 & 2430.21 \\
 \hline
 Inspiral 50 & 4341.32 & 4341.56 & 4334.59 & 4334.95 \\
 \hline
 Inspiral 100 & 8061.46 & 8061.69 & 7301.31 & 7301.91 \\
 \hline
 Inspiral 1000 & 76308.05 & 76308.28 & 76033.17 & 76038.57 \\
 \hline
 Montage 25 & 99.13 & 99.38 & 91.02 & 91.17 \\
 \hline
 Montage 50 & 198.53 & 198.84 & 187.76 & 187.96 \\
 \hline
 Montage 100 & 393.73 & 394.08 & 392.5 & 392.79 \\
 \hline
 Montage 1000 & 4022.71 & 4024.32 & 4021.67 & 4023.74 \\
 \hline
 Sipht 30 & 5396.21 & 5396.66 & 4460.96 & 4461.62 \\
 \hline
 Sipht 60 & 10443.98 & 10444.39 & 5008.16 & 5008.54 \\
 \hline
 Sipht 100 & 10874.64 & 10874.85 & 6204.36 & 6200.89 \\
 \hline
 Sipht 1000 & 72895.54 & 72895.65 & 57894.87 & 57897.12 \\
 \hline
\end{tabular}
\end{adjustbox}
\caption{Execution time of WorkflowSim and WIDESim for all scientific workflows \label{table:compute_results}}
\end{table}

\subsubsection{Network model evaluation}

\begin{figure}[h]
\centering
\includegraphics[width=0.5\textwidth]{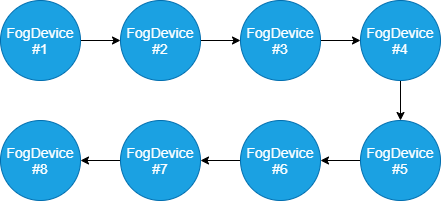}
\caption{The network topology used in comparing iFogSim vs WIDESim}
\label{fig:network_net_model}
\end{figure}

We assess the validity of WIDESim in terms of the network model by comparing its performance with iFogSim. According to \cite{gupta2017ifogsim}, iFogSim only supports tree-like topologies. Therefore we have to evaluate the network model of the simulators using a more restricted topology compared with WIDESim. Another problem is the way messages are sent between different devices in iFogSim. Although in iFogSim, each node broadcasts its message on its outgoing links, WIDESim uses routing tables to propagate messages in the network. So, it is expected to experience various congestion and delays in the performance of these two simulators over the same network topology. However, a linear topology prevents this difference, so we choose the network topology of figure \ref{fig:network_net_model} in this comparison. \begin{figure}[ht]
\centering
\includegraphics[width=0.5\textwidth]{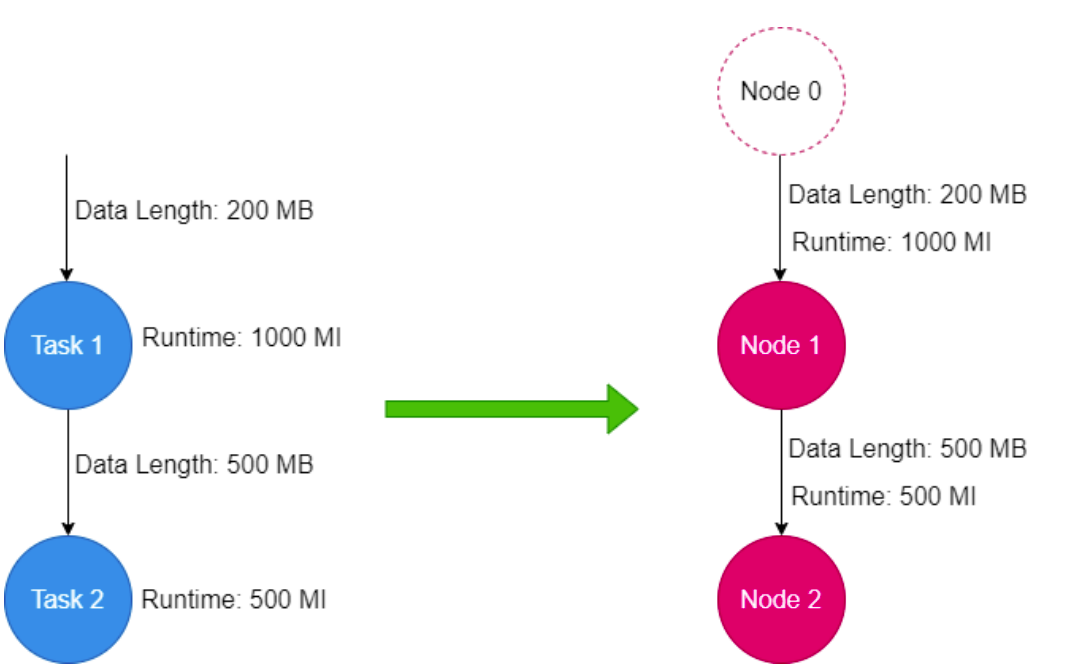}
\caption{Successful mapping of iFogSim application to WIDESim workflow}
\label{fig:ifog_sim_conersion_1}
\end{figure}

\begin{figure}[ht]
\centering
\includegraphics[width=0.5\textwidth]{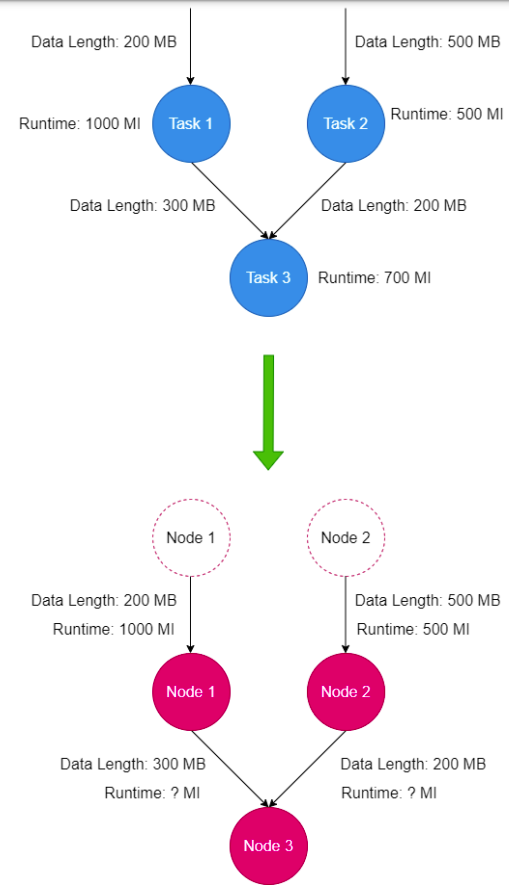}
\caption{Unsuccessful mapping of iFogSim application to WIDESim workflow}
\label{fig:ifog_sim_conersion_2}
\end{figure}

In addition to the differences mentioned in the implementation of the network model of these two simulators, the computational model of these two also has differences. The input of the computational model of both simulators is a DAG, but the meaning of node and edge in in the computational model of the iFogSim is different from WIDESim. In a workflow in WIDESim, a node represents a task, and each task has an input file, an execution time, and an output file. This task must be scheduled and prepared for execution on a virtual machine. Also, each edge represents a data dependency between tasks. The weight of each edge also indicates the amount of data that the source task produces and the destination task consumes. On the other hand, in the iFogSim computing model, each node represents a virtual machine which only one application module is assigned to it, unlike WIDESim where each vm can schedule multiple tasks. Each edge contains information that informs the destination virtual machine how much data it should receive and how many instructions it should execute. This means that a one-to-one conversion from a workflow to an application is not necessarily possible in the iFogSim.

In scenarios with linear workflow, this conversion is one-to-one, and as shown in Figure \ref{fig:ifog_sim_conersion_1}, the amount of data required to complete each intermediate task in the workflow must be placed on the input edge of that task in a similar node in the iFogSim's computational model. But as shown in Figure \ref{fig:ifog_sim_conersion_2}, there are scenarios in which the workflow model cannot provide enough information for the computational model of iFogSim and therefore the conversion fails. In this figure, the reason for this failure is that Task 3 has two parents, and this makes us unable to determine which edge to allocate the necessary execution time to in conversion. A solution to this problem is to divide the task execution time between the input edges in a weighted manner, means that execution time is assigned based on data length which is transferred on each edge. But for several reasons, this method may lead to the implementation of the application being different from the workflow that was generated from it as is mentioned below.

\begin{itemize}
  \item In the case of workflow, it is meant that Task 3 is done at once, but in the conversion of this task into two tasks as shown in Figure \ref{fig:ifog_sim_conersion_2}, depending on the scheduling algorithm in Node 3, different results can be generated. In other words, the scheduling algorithm of Node 3 may schedule whole 700 instructions in a different way that schedules two smaller tasks and it causes inconsistency in two simulator's results comparison.
  \item Also, another case is that Node 3 executes two received tasks when both tasks have been sent. For example, if Node 1 finishes its execution and sends its output to Node 3 along with the number of commands necessary to process it, Node 3 must wait for the output file of Node 2 to be generated and reach it. Because Task 3 executes all 700 million commands at once. But since it is not possible to determine such a property in this simulator, the execution of the application will be different from the reference workflow.
\end{itemize}

\begin{table}
\centering

\begin{tabular}{ |c|c| } 
 \hline
 \textbf{Parameter} & \textbf{Value} \\ 
  \hline
 Number of VMs & 8 \\ 
 \hline
 VM size & 10000 MB \\
 \hline
 RAM & 512 MB \\ 
 \hline
 Bandwidth & 1024 MB/s \\
 \hline
 MIPS & 1000 MIPS \\ 
 \hline
 Number of processing cores & 1 \\
 \hline
 Scheduling Algorithm & CloudletSchedulerTimeShared \\ 
 \hline
 Number of Datacenters & 8 \\
 \hline
 Number of Hosts & 1 \\ 
 \hline
 Number of Processing Cores for Each Host & 1 \\
 \hline
 MIPS of Each Host & 1000 MIPS \\ 
 \hline
 Processing Core Provisioning Algorithm & PeProvisionerSimple \\
 \hline
 RAM of Each Host & 512 MB \\ 
 \hline
 RAM Provisioning Algorithm & RamProvisionerSimple \\
 \hline
 Bandwidth of Each Host & 1024 MB/s \\
 \hline
 Bandwidth Provisioning Algorithm & BwProvisionerSimple \\ 
 \hline
 VM Allocation Policy & VmAllocationPolicySimple \\
 \hline
 VM Scheduler Algorithm & VmSchedulerTimeShared \\
 \hline
\end{tabular}
\caption{Specification of the simulation environment in evaluating WIDESim vs iFogSim\label{table:network_net_spec}}
\end{table}

\begin{figure}[ht]
\centering
\includegraphics[width=0.7\textwidth]{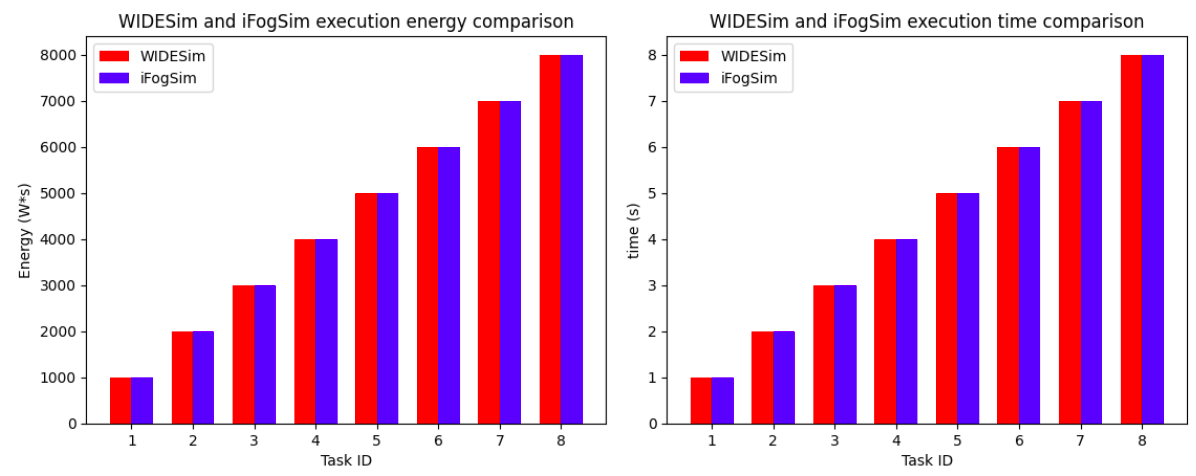}
\caption{WIDESim execution time and energy compared to iFogSim}
\label{fig:ifog_sim_result}
\end{figure}

Figure \ref{fig:ifog_sim_result} shows the comparison of the performance of WIDESim with iFogSim in terms of execution time and energy consumption. We see the same performance of WIDESim and iFogSim under the same topology and simulate the same application. This result is, in conjunction with the results obtained by comparing WIDESim to WorkflowSim, showing WIDESim can be used for simulating different types of scientific workflows over different network topologies (graph topology), with close performance to standard simulators that are an extension of CloudSim.

\subsection{Analyzing the contributions of WIDESim}

In this section, we are going to evaluate the advanced features of WIDESim. These features include supporting complex graph topologies and a variety of workflows, including single, multiple, and workflow ensembles. 

In the assessments of this section, we've mainly focused on two essential factors: time and energy consumption. These are crucial aspects that require efficient management, especially in cloud, fog, and edge computing. For all our experiments, we've consistently used the FCFS algorithm for the assignment of tasks to virtual machines. Additionally, our internal virtual machine scheduling algorithms are designed to share processing time efficiently.

\subsubsection{Case studies} \label{sec:topologies}
\label{case}
\begin{figure}[ht]
\centering
\includegraphics[width=0.5\textwidth]{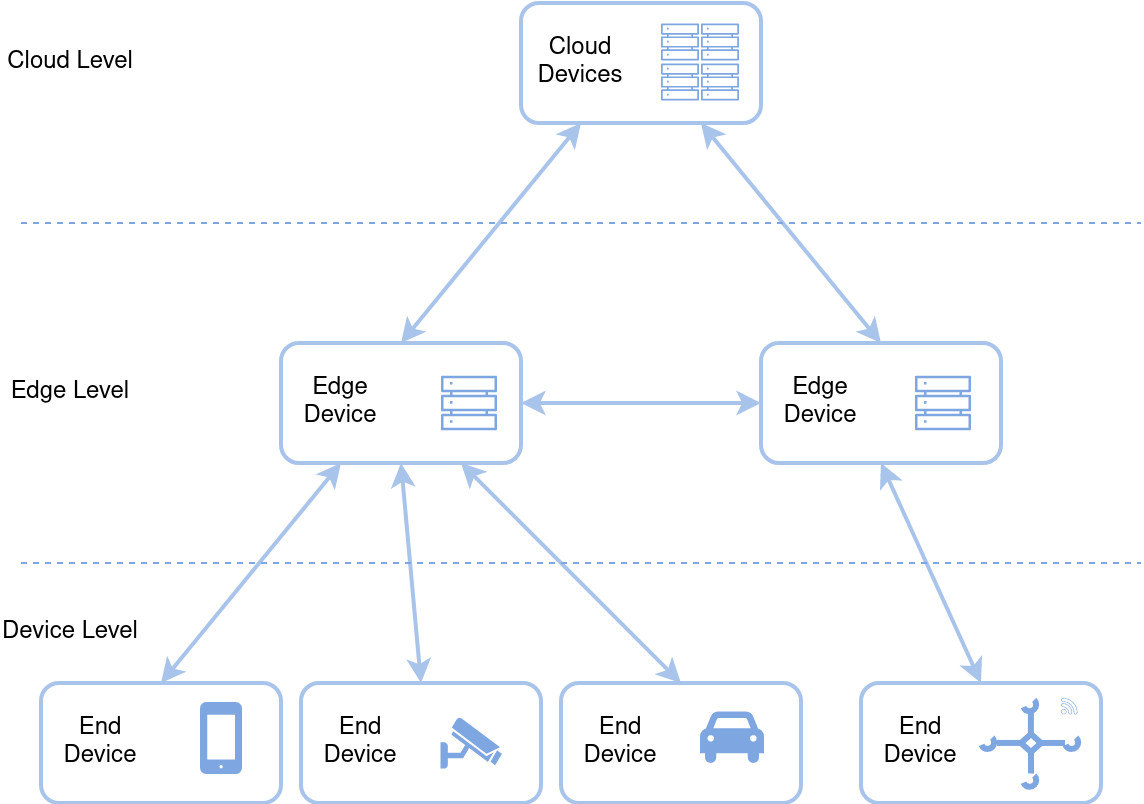}
\caption{Edge server topology}
\label{fig:topology-a}
\end{figure}

In this part, we will study the performance of WIDESim in three different network topologies, which are presented in 
\cite{tocze2018taxonomy}. The first topology is called edge server, in which cloud devices are connected to the edge 
devices and edge devices are connected to end devices. This topology illustrates edge computing architecture and cloudlet, where bandwidth and computationally intensive tasks are sometimes expected. In this topology, thanks to edge devices, it is possible to manage bandwidth challenges in cloud computing for bandwidth-intensive applications. Edge devices, with less computational power than cloud devices, can also process all or some of the tasks of the applications. Relatively high computational power, moderate power consumption, and lower latency are the advantages of this topology. This topology is shown in figure \ref{fig:topology-a}.

\begin{figure}[ht]
\centering
\includegraphics[width=0.5\textwidth]{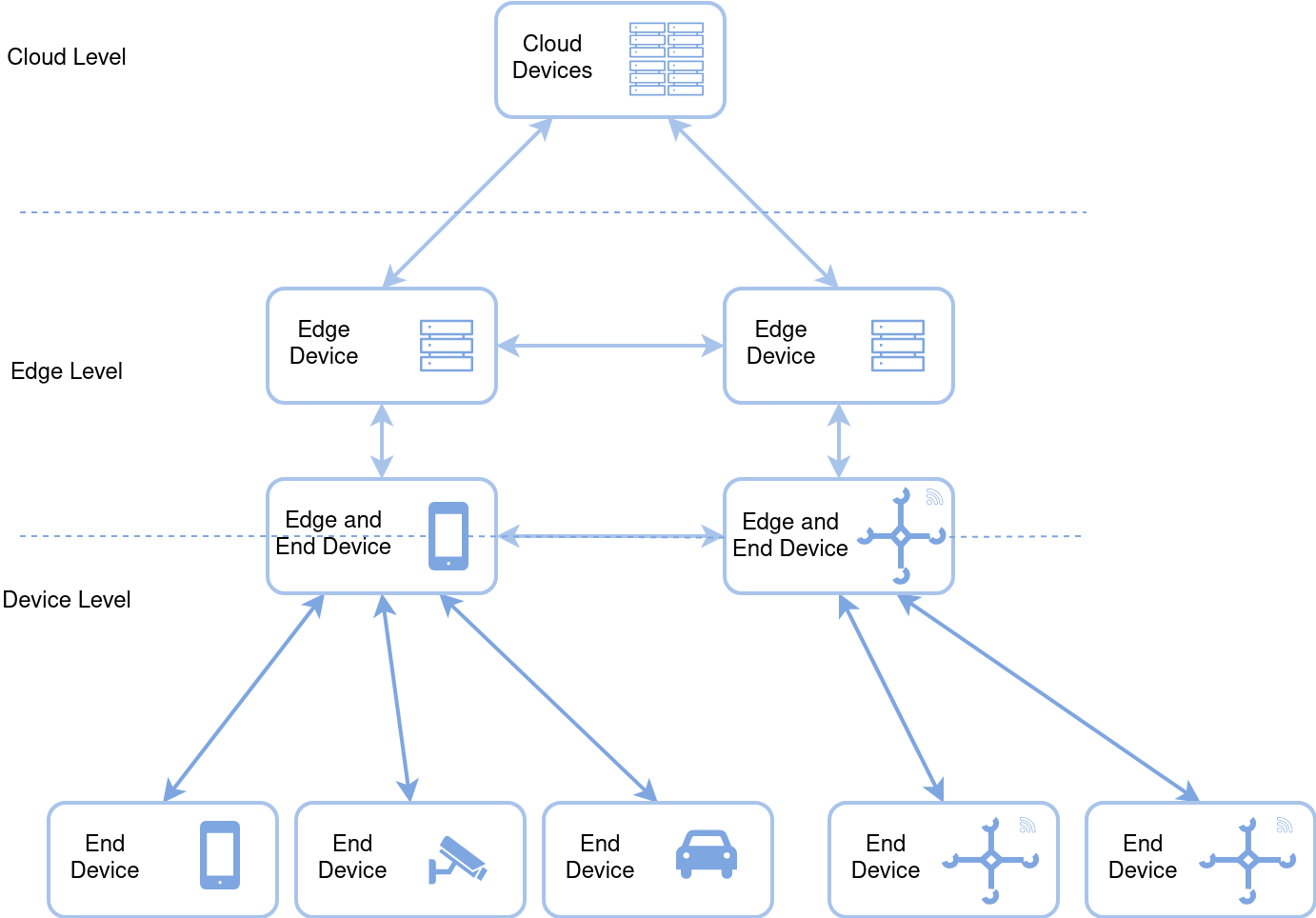}
\caption{Coordinator device topology}
\label{fig:topology-b}
\end{figure}

The second topology is called coordinator device, in which both edge and end devices are considered computational resources, and these edge devices are called coordinators. In this topology, some end devices can mediate between the cloud and other end devices. In other words, the coordinator 
devices are connected to edge devices and other end devices. There is not any specific separation between the edge and end layers. Even the edge layer can contain a hierarchy of devices.
In this topology, fog and end devices are present in the range of end-to-cloud devices. Hence, this topology is suitable for both computational power-intensive and bandwidth-intensive tasks since end devices can do some computations without needing cloud devices. 
However, running a computationally intensive workflow may result in an increase in the consumed power. Since fog and edge devices have higher bandwidth than cloud devices, this topology can decrease the transmission time of tasks and data. This topology is shown in figure \ref{fig:topology-b}.

\begin{figure}[ht]
\centering
\includegraphics[width=0.5\textwidth]{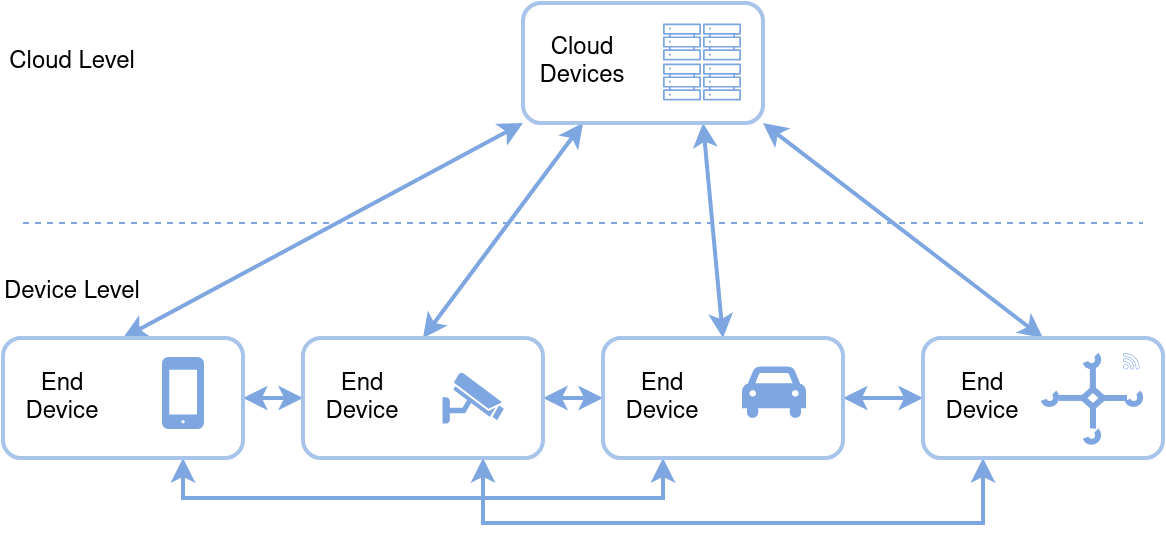}
\caption{Device cloud topology}
\label{fig:topology-c}
\end{figure}

The third topology is called device cloud. Unlike previously mentioned ones, this topology has less interaction between the end and cloud devices. End devices can also act as small cloud devices and do
some of the cloud devices' tasks. The more computationally intensive tasks are still done by cloud devices. This topology can illustrate the mobile-cloud computing environment, especially in the case of decentralized and peer-to-peer communication. Since fog and edge devices are computationally less capable, and cloud devices are less used in this topology, this topology is unsuitable for massive workloads with computationally intensive tasks. Power consumption and latency are expected to vary depending on the network configuration. This topology is shown in figure \ref{fig:topology-c}.

These topologies can be seen in a wide range of IoT applications.
For the first example, we can consider a transportation scenario in which smart cars, traffic lights, and other sensors should 
be connected to each other. In this scenario, the first topology can be used.
For the second example, consider visual security and surveillance scenarios in which cameras are deployed in different places. 
These cameras can generate massive amounts of data. Traditionally, these cameras are connected to a central cloud server.
But, nowadays, by increasing camera resolution and frame rate, traditional cloud-based architectures are not efficient and scalable.
So, we can use the second topology for this scenario.
As the last example, we can consider a smart building scenario in which different sensors are deployed in a building. These sensors
can be used for different purposes like security, energy management, etc. In this scenario, we can use the third topology.
\cite{tocze2018taxonomy}

The simulation settings, common to all three topologies, are detailed in Table \ref{table:case_topo_spec}. The detailed specification of devices of each topology (edge server, coordinator server, and device cloud) can also be found in tables \ref{table:case_a_topo_spec}, \ref{table:case_b_topo_spec}, and \ref{table:case_c_topo_spec} respectively.

\begin{table}
\centering

\begin{tabular}{ |c|c| } 
 \hline
 \textbf{Parameter} & \textbf{Value} \\ 
 \hline
 VM size & 10000 MB \\
 \hline
 Scheduling algorithm & CloudletSchedulerTimeShared \\ 
 \hline
 Processing core Provisioning algorithm & PeProvisionerSimple \\
 \hline
 RAM Provisioning algorithm & RamProvisionerSimple \\
 \hline
 Bandwidth Provisioning algorithm & BwProvisionerSimple \\ 
 \hline
 VM allocation policy & VmAllocationPolicySimple \\
 \hline
 VM scheduler algorithm & VmSchedulerTimeShared \\
 \hline
 Task to VM mapping algorithm & FCFS \\
 \hline
\end{tabular}
\caption{General specification of all topologies in evaluating WIDESim\label{table:case_topo_spec}}
\end{table}

\begin{table}
\centering

\begin{adjustbox}{width=1\textwidth} \begin{tabular}{ |c|c|c|c|c| } 
 \hline
 \textbf{Parameter} & \textbf{Cloud Device} & \textbf{Edge Device} & \textbf{\makecell{End Device\\(Phone)}} & \textbf{\makecell{End Device\\(Other)}} \\ 
 \hline
 Number of VMs & 1 & 1 & 1 & 1 \\ 
 \hline
 RAM of VM & 1024 MB & 512 MB & 64 MB & 64 MB \\ 
 \hline
 Bandwidth of VM & 512 MB/s & 1000 MB/s & 4000 MB/s & 4000 MB/s \\
 \hline
 MIPS of VM & 6000 MIPS & 4000 MIPS & 500 MIPS & 500 MIPS \\ 
 \hline
 Number of VM processing cores & 1 & 1 & 1 & 1 \\
  \hline
 Number of devices & 1 & 2 & 1 & 3 \\
  \hline
 Number of hosts per device & 1 & 1 & 1 & 1 \\ 
 \hline
 \makecell{Number of processing cores \\ for each host} & 1 & 1 & 2 & 1 \\
  \hline
 MIPS of each host & 6000 MIPS & 4000 MIPS & 500 MIPS & 500 MIPS \\ 
  \hline
 RAM of each host & 1024 MB & 512 MB & 64 MB & 64 MB \\
  \hline
 Bandwidth of each host & 512 MB/s & 1000 MB/s & 4000 MB/s & 4000 MB/s \\
 \hline
\end{tabular}
\end{adjustbox}
\caption{Device specification of topology edge server\label{table:case_a_topo_spec}}
\end{table}

\begin{table}
\centering

\begin{adjustbox}{width=1\textwidth} \begin{tabular}{ |c|c|c|c|c|c| } 
 \hline
 \textbf{Parameter} & \textbf{Cloud Device} & \textbf{Edge Device} & \textbf{\makecell{Edge and \\ End Device}} & \textbf{\makecell{End Device\\(Phone)}} & \textbf{\makecell{End Device\\(Other)}} \\ 
 \hline
 Number of VMs & 1 & 1 & 1 & 1 & 1 \\ 
 \hline
 RAM of VM & 1024 MB & 512 MB & 64 MB & 64 MB & 64 MB \\ 
 \hline
 Bandwidth of VM & 512 MB/s & 1000 MB/s & 5000 MB/s & 5000 MB/s & 5000 MB/s \\
 \hline
 MIPS of VM & 6000 MIPS & 4000 MIPS & 1500 MIPS & 500 MIPS & 500 MIPS \\ 
 \hline
 \makecell{Number of VM \\ processing cores} & 1 & 1 & 1 & 1 & 1 \\
  \hline
 Number of devices & 1 & 2 & 2 & 1 & 4 \\
  \hline
 Number of hosts per device & 1 & 1 & 1 & 1 & 1 \\ 
 \hline
 \makecell{Number of processing cores \\ for each host} & 1 & 1 & 1 & 2 & 1 \\
  \hline
 MIPS of each host & 6000 MIPS & 4000 MIPS & 1500 MIPS & 500 MIPS & 500 MIPS \\ 
  \hline
 RAM of each host & 1024 MB & 512 MB & 64 MB & 64 MB & 64 MB \\
  \hline
 Bandwidth of each host & 512 MB/s & 1000 MB/s & 5000 MB/s & 5000 MB/s & 5000 MB/s \\
 \hline
\end{tabular}
\end{adjustbox}
\caption{Device specification of topology coordinator device\label{table:case_b_topo_spec}}
\end{table}

\begin{table}
\centering

\begin{tabular}{ |c|c|c|c| } 
 \hline
 \textbf{Parameter} & \textbf{Cloud Device} & \textbf{\makecell{End Device \\ (Phone)}} & \textbf{\makecell{End Device \\ (Other)}} \\ 
 \hline
 Number of VMs & 1 & 1 & 1 \\ 
 \hline
 RAM of VM & 1024 MB & 64 MB & 64 MB \\ 
 \hline
 Bandwidth of VM & 512 MB/s & 4500 MB/s & 4500 MB/s \\
 \hline
 MIPS of VM & 6000 MIPS & 500 MIPS & 500 MIPS \\ 
 \hline
 Number of VM processing cores & 1 & 1 & 1 \\
  \hline
 Number of devices & 1 & 1 & 3 \\
  \hline
 Number of hosts per device & 1 & 1 & 1 \\ 
 \hline
 Number of processing cores for each host & 1 & 2 & 1 \\
  \hline
 MIPS of each host & 6000 MIPS & 500 MIPS & 500 MIPS \\ 
  \hline
 RAM of each host & 1024 MB & 64 MB & 64 MB \\
  \hline
 Bandwidth of each host & 512 MB/s & 4500 MB/s & 4500 MB/s \\
 \hline
\end{tabular}
\caption{Device specification of topology device cloud\label{table:case_c_topo_spec}}
\end{table}

\subsubsection{Single Workflow Evaluation}
\label{singleeval}

In our evaluations thus far, we have constrained WIDESim's capabilities to ensure equal conditions for conducting experiments and making fair comparisons with other simulators. These limitations made us restrict the network topologies and workflows used in our experiments. As a result, we couldn't fully showcase WIDESim's capabilities and thoroughly assess its ability to handle complex workflows and network scenarios.

Recognizing these limitations, this section focuses on understanding WIDESim's strengths and validating its network and computational abilities. The workflows we've used for this assessment are the same scientific workflows we used in previous evaluations when comparing WIDESim with WorkflowSim. In each test, we've simulated one of these workflows in different computing environments. Regarding the network topologies we've used, we've chosen three practical configurations: edge server, coordinator device, and device cloud, as we discussed earlier. These topologies have more complex structures, including loops and cycles, which are challenging for other simulators to handle. These topologies are inspired by real-world networks and demonstrate WIDESim's ability to simulate real-world scenarios accurately.

\begin{figure}
\centering
\begin{minipage}{0.31\textwidth}
\includegraphics[width=\linewidth]{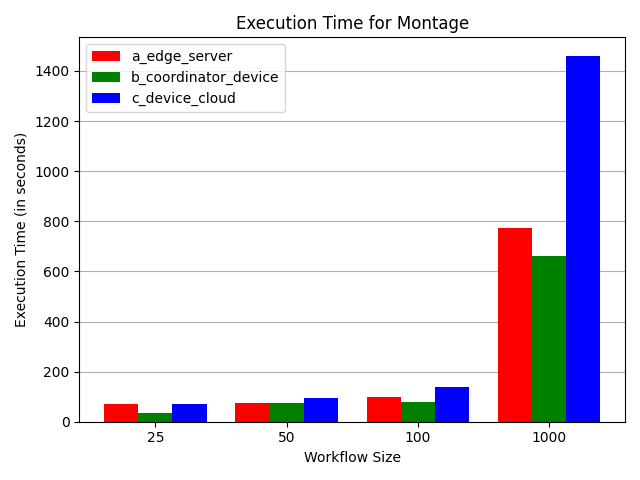}
\subcaption{Montage}
         \label{fig:montage_single_time_result}
\end{minipage}
\hspace{1mm} 
\begin{minipage}{0.31\textwidth}
\includegraphics[width=\linewidth]{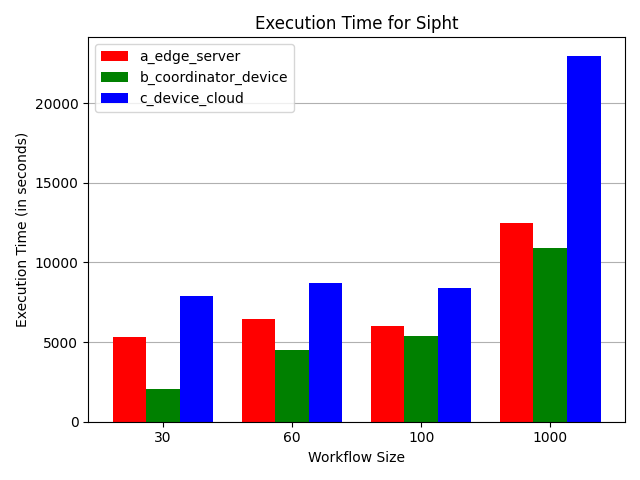}
\subcaption{Sipht}
         \label{fig:sipht_single_time_result}
\end{minipage}

\medskip
\begin{minipage}{0.31\textwidth}
\includegraphics[width=\linewidth]{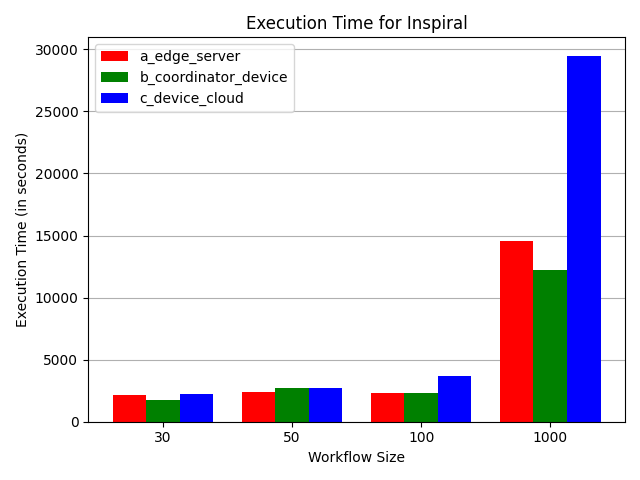}
\subcaption{Inspiral}
         \label{fig:inspiral_single_time_result}
\end{minipage}
\hspace*{\fill}
\begin{minipage}{0.31\textwidth}
\includegraphics[width=\linewidth]{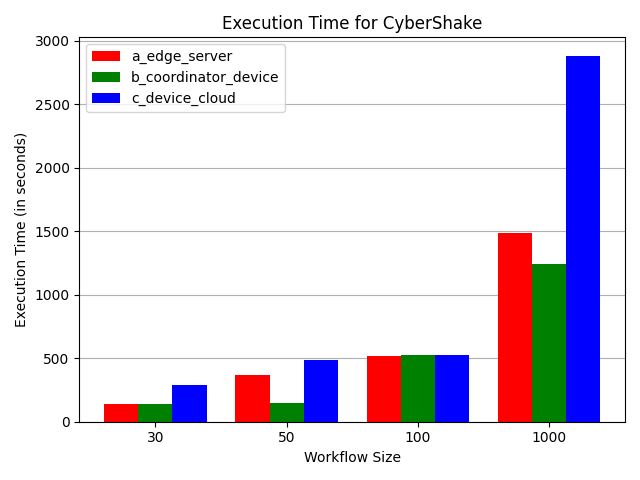}
\subcaption{Cybershake} \label{fig:cybershake_single_time_result}
\end{minipage}
\hspace*{\fill}
\begin{minipage}{0.31\textwidth}
\includegraphics[width=\linewidth]{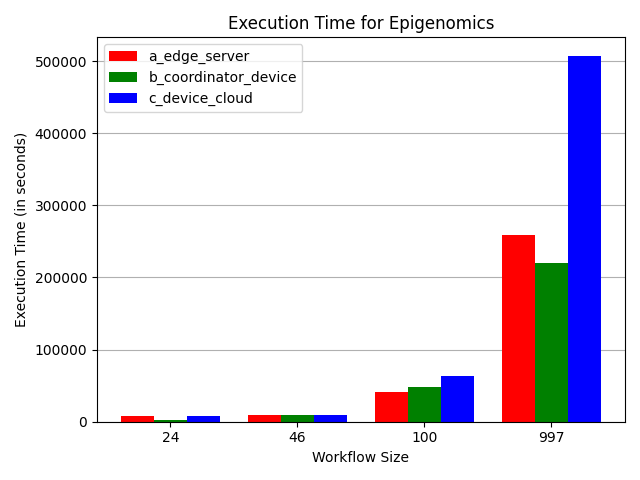}
\subcaption{Epigenomics} \label{fig:epigenomics_single_time_result}
\end{minipage}

\caption{Execution time of individual scientific workflows on three different topologies in the WIDESim}
\label{fig:single_workflow_time_results}
\end{figure}

\begin{figure}
\centering
\begin{minipage}{0.31\textwidth}
\includegraphics[width=\linewidth]{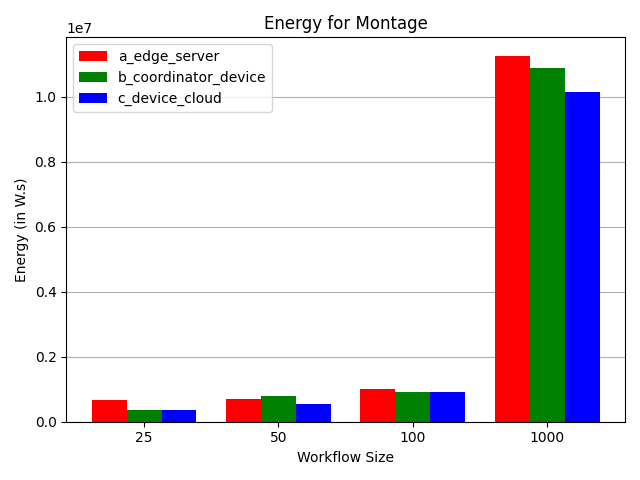}
\subcaption{Montage}
         \label{fig:montage_single_energy_result}
\end{minipage}
\hspace{1mm} 
\begin{minipage}{0.31\textwidth}
\includegraphics[width=\linewidth]{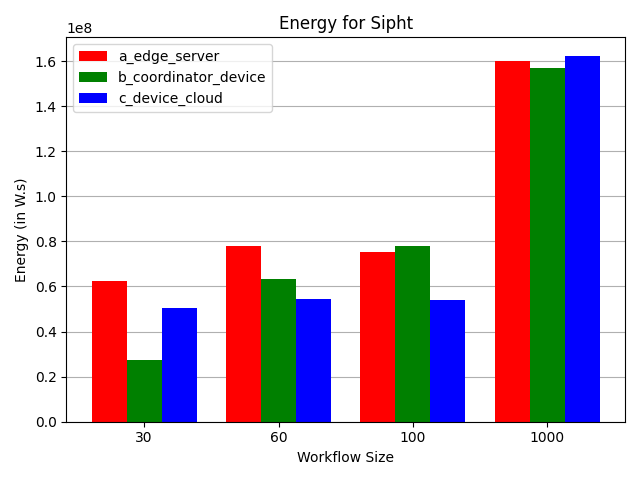}
\subcaption{Sipht}
         \label{fig:sipht_single_energy_result}
\end{minipage}

\medskip
\begin{minipage}{0.31\textwidth}
\includegraphics[width=\linewidth]{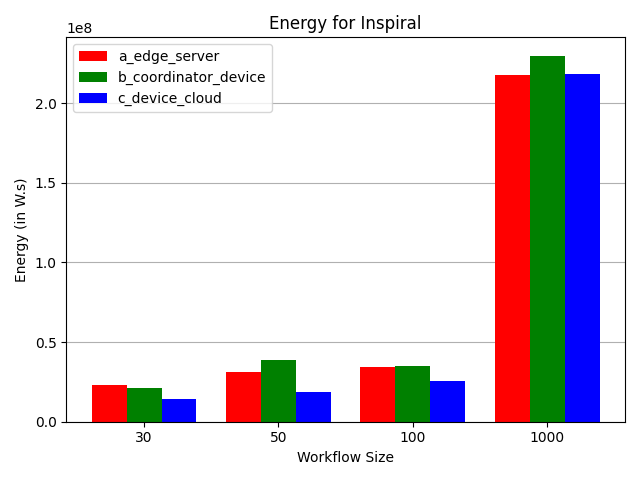}
\subcaption{Inspiral}
         \label{fig:inspiral_single_energy_result}
\end{minipage}
\hspace*{\fill}
\begin{minipage}{0.31\textwidth}
\includegraphics[width=\linewidth]{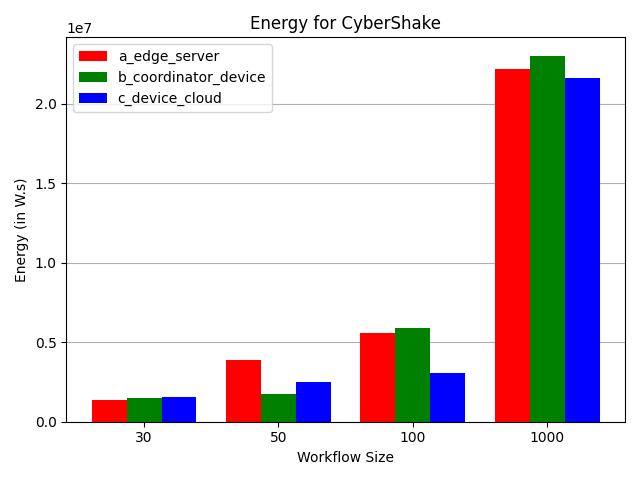}
\subcaption{Cybershake} \label{fig:cybershake_single_energy_result}
\end{minipage}
\hspace*{\fill}
\begin{minipage}{0.31\textwidth}
\includegraphics[width=\linewidth]{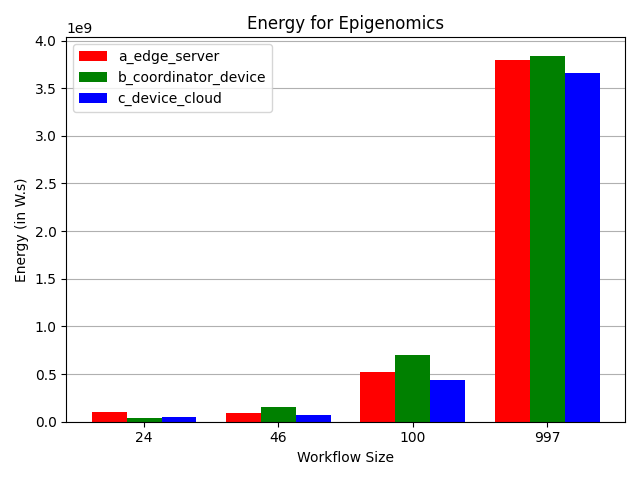}
\subcaption{Epigenomics} \label{fig:epigenomics_single_energy_result}
\end{minipage}

\caption{Energy consumption of individual scientific workflows on three different topologies in the WIDESim}
\label{fig:single_workflow_energy_results}
\end{figure}

To enable a comparison of the execution results of individual workflows on different topologies, we've created bar charts. In these charts, each distinct workflow size is represented with different colors for each of the three reference realistic topologies. Figure \ref{fig:single_workflow_time_results} displays the charts related to workflow execution time, and Figure \ref{fig:single_workflow_energy_results} shows those related to energy consumption. All execution results are available in the directory \href{https://github.com/ARH80/WIDESim/tree/master/outputs/single_workflow}{\texttt{outputs/single\_workflow}} within the project's GitHub repository. The results are categorized by network topology in this directory. In these files, you can find the results related to energy consumption at the end of the simulation run and above the summary table for task execution times.

Regarding the results obtained from the evaluation of individual workflows, we can observe that in almost all cases, as the workflow size increases, the time required for processing all tasks follows an increasing trend in all three topologies. This upward trend in energy consumption is also evident with an increase in the workflow size. In some cases, such as the Sipht workflow, the upward trend is less apparent for smaller sizes. This phenomenon is due to the efficient concurrent execution structure within these workflows. Essentially, due to lower workloads and greater potential for concurrent execution among tasks in these workflows, the time difference between executions is reduced. These findings align with expectations, as devices take more time to complete all tasks with an increase in the number of executable tasks. Since energy consumption is dependent not only on device power but also on simulation runtime, it similarly increases with execution time.

Another interesting aspect to investigate is the general comparison of execution time and energy consumption in different topologies, independent of the processed workflow. By examining the execution time results for different workflows individually on these three topologies, we observe that, in most cases, the execution time for a workflow on the device cloud topology (c) is significantly higher than the other two topologies. These observations are justified considering the differences among these three topologies. As evident from the device cloud (c) topology's structure, it essentially represents a simplified cloud environment, with most computations dedicated to a specific cloud data-center. This is in contrast to the other two topologies that focus more on edge and fog environment structures. Therefore, for executing a set of tasks on the device cloud topology (c), a single device bears the majority of computations, resulting in an extended execution time. Additionally, considering that in all links between edge and data-center devices, the lower bandwidth of cloud device is the bottleneck of communication, file transfer times between devices are significantly increased in this topology. This is while in the other topologies, the participation of various edge and fog devices in the environment and their collaboration with the cloud server enhances the overall processing power, requiring less time to complete all tasks in the environment.

On the other hand, when comparing the execution times of workflows in the edge server topology (a) to the coordinator device topology (b), considering the architecture of these topologies is of great importance. The edge server topology, in addition to the end devices, includes edge servers near the end users with relatively high processing power, allowing a larger portion of computations to be performed outside the cloud device, within the edge layer. This topology, overall, offers higher processing capabilities compared to the device cloud topology (c) and, due to its emphasis on task processing at edge devices, represents an edge computing environment. The coordinator device topology (b) also places a significant emphasis on the processing power available at the edge. It includes end devices with slightly higher processing power connected to the regular end devices, which can function as less powerful edge servers, referred to as edge-end devices. These edge-end devices are connected to more powerful edge servers in higher layers, still near end users, enabling the processing of heavier workloads within the edge environment. The higher number of employed edge and end devices and the diversity of their interconnections unveil a substantial amount of readily accessible computational resources in the space between the cloud device and end devices. The intricate hierarchical and widely distributed architecture of the devices in the network makes the coordinator device topology (b) more representative of a fog computing environment.

Considering the architectures of these two topologies, the overall processing power of the coordinator device topology (b), particularly in the edge layer, is greater than that of the edge server topology (a). As a result, we generally expect the execution time for a set of similar tasks in the coordinator device topology (b) to be less than that in the edge server topology (a). Based on the results obtained in Figure \ref{fig:single_workflow_time_results}, we see that this pattern has been followed in almost all workflow cases, with the coordinator device topology (b) having the shortest execution time among all three topologies. In very few cases, such as Inspiral50 and Epigenomics100, a slightly larger execution time is observed in the coordinator device topology (b) compared to the edge server topology(a). The reason for this observation is that the evaluation results exhibit a strong dependence on various factors such as workflow graph structure, task size, task dependencies, number of devices and interconnections in topology, and task mapping and scheduling algorithms optimality, rather than just the overall processing power of the environment. The impact of the processing power shows itself more strongly in larger number of tasks. The FCFS algorithm, for instance, makes task mapping decisions solely based on a simple heuristic of task arrival time, without considering critical parameters like device processing power, bandwidth of connections, task size, and task dependencies. Therefore, as expected, the outcome of mapping and scheduling with this algorithm may not be optimal, leading to greater delays and suboptimal processing times in some cases. The few instances where the general execution time trend between the edge server (a) and coordinator device (b) topologies has deviated slightly are influenced by this phenomenon.

Looking at energy consumption in the results obtained from the evaluation, we observe that for each single workflow type scenario, energy consumption for smaller sizes is lower in the device cloud topology (c) compared to the edge server (a) and coordinator device (b) topologies. However, for larger workflow sizes, this observation is inverted, and energy consumption in the device cloud topology (c) surpasses the other two topologies. The reason behind this lies in how we calculate the energy consumption in computational environments. Energy consumption is directly related to both the processing power of devices and the total runtime of workflow execution. Therefore, for smaller workflow sizes, despite the slightly longer execution time of the device cloud topology (c) compared to the other two topologies, that's mainly because of the large file transfer times due to limited bandwidth in the cloud layer and poor utilization of devices in the edge layer, the lower power consumption of the devices in this topology (primarily consisting of the power consumption of a single cloud device) dominate the calculation of energy consumption. As a result, in this scenario, the device cloud topology (c) consumes less energy compared to the other two topologies. In cases where the workflow size is larger, and the number of tasks is higher, due to network congestion in the device cloud topology (c) and its limited bandwidth, the execution time significantly outweighs the power consumption of devices in the environment, which remains constant. Consequently, in this scenario, the device cloud topology (c) consumes more energy compared to the other two topologies.

Furthermore, as previously mentioned, the coordinator device topology (b) has more devices and interconnections in its network than the edge server topology (a).This leads to the expectation that, despite generally having a shorter execution time compared to the edge server topology (a), the coordinator device topology (b) usually consumes more energy under similar conditions than the edge server topology (a). Observing the energy consumption results for different topologies in Figure \ref{fig:single_workflow_energy_results} also confirms this general expectation to a considerable extent. In some cases where the opposite pattern is observed, and the energy consumption in the coordinator device topology (b) is lower than that in the edge server topology (a), considering the small total number of tasks, as well as the workflow graph structure and the size of the executable tasks, becomes a key factor. For example, in the Cybershake50 workflow where the energy consumption in the coordinator device topology (b) is significantly lower than that in the edge server topology (a), it can be attributed to the small task size and the need for less execution time. Most of the workload is handled by the end devices in the low-power edge layer, leading to a significant reduction in the overall energy consumption of this topology. This is due to the higher number and diversity of end devices in the coordinator device topology (b) compared to the edge server topology (a).

\subsubsection{Multiple Workflow Evaluation}

One of the most important features of the WIDESim simulator is support for the concurrent execution of several different workflows. To perform this, according to the paper \cite{goudarzi2020application}, a list of workflows is passed to the TaskManager class, which is responsible for managing tasks. Then, depending on the scheduling algorithm specified by the user, the tasks are scheduled. It is also crucial to note that in the implementation of the Task class in the WIDESim simulator, an identifier related to the respective workflow is stored for each task, indicating which workflow each task belongs to. This feature in the WIDESim simulator provides the possibility to investigate the performance and execution method of each of the workflows separately.

\begin{figure}[ht]
    \centering
    \includegraphics[width=.8\textwidth]{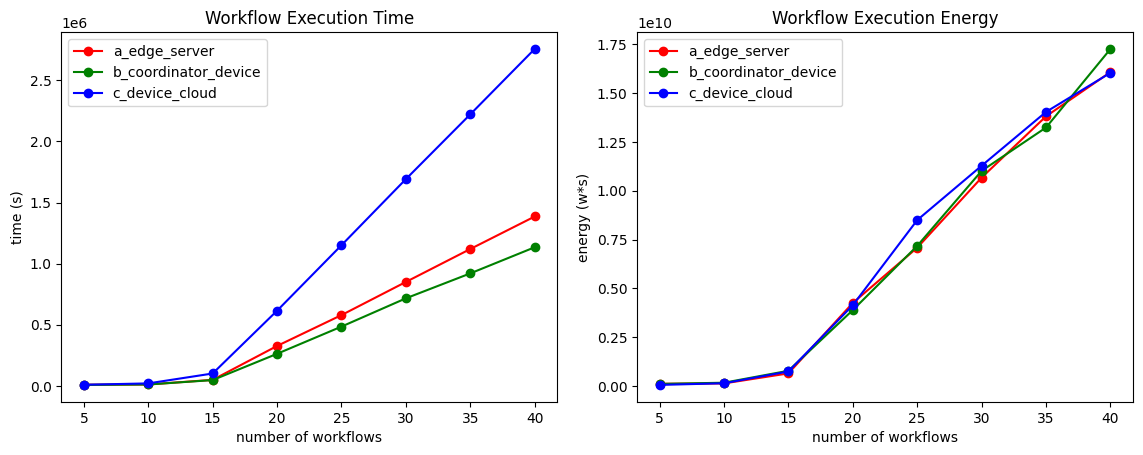}
    \caption{Duration and Energy consumption of multi-workflow evaluation scenarios on different topologies}
    \label{fig:topology-evalmulti}
\end{figure}

In the evaluation phase of this WIDESim simulator feature, the network topologies mentioned in the case study section \ref{sec:topologies} have been used, and benchmark workflow groups have been used as workflows. To estimate and examine this matter more precisely, evaluations have been performed on eight scenarios of multiple workflows where there are 5, 10, 15, 20, 25, 30, 35, and 40 workflows. In the scenario with 5 workflows the smallest size of each five benchmark workflows in table \ref{table:compute_results} is used for simulation. For further evaluations, as the number of workflows increases, the set of 5 benchmark workflows with larger size is added to simulation environment. In each scenario, one instance of each of the 5 larger benchmark workflows is added to the workflow groups until the largest size of each 5 benchmark workflows, afterwards the largest size of benchmark workflows are used. The exact structure of workflows used and the simulation code in each scenario can be viewed at GitHub repository.
We evaluate each of the 8 scenarios of multiple workflows on the three mentioned network topologies, and we compare the results in terms of execution time and energy in each. The time and energy results related to different topologies are available in figure \ref{fig:topology-evalmulti}. Also, detailed and precise reports related to each simulation, such as information and detailed data for each task and other data, are accessible at the path \href{https://github.com/ARH80/WIDESim/tree/master/outputs/multiple_workflows}{\texttt{outputs/multiple\_workflows}} in the GitHub repository.

According to the mentioned diagrams, during the increase in the number of workflows initially, due to the fact that the dimensions of each workflow are increasing in addition to their number, the slope of time and energy diagrams is smaller than the case where the dimensions of workflows grow to a fixed size at the end of the x-axis.
The increase, as expected, in the amount of time and energy during the increase in the number and size of workflows confirms the simulator's accuracy in this regard. In terms of comparing the network topologies, the execution time in device cloud scenario is much more than other two topologies due to the use of only cloud devices for processing, unlike other scenarios where edge and end devices are also used as processing units, another reason is the network latency which is larger in the case of cloud scenario comparing other two topologies. Moreover, the execution time of edge server scenario is more than coordinator device, which is consistent with the results in single execution scenario in figure \ref{fig:single_workflow_time_results} that is mainly due to more processing power in coordinator device topology.

When evaluating energy consumption in single workflow scenarios, it was observed that edge and fog setups tend to consume more energy compared to cloud configurations. This is illustrated in Figure \ref{fig:single_workflow_energy_results}, where, in most instances, the energy usage of edge servers and coordinator devices exceeds that of cloud-based devices. However, considering a group of benchmark workflows with different structures doesn't lead to a same result as single workflow evaluation. Thus, energy consumption of multiple workflows does not follow a specific pattern as it does in single workflow scenario due to different structure of combined workflows and non-optimized scheduling and allocation algorithms.

\begin{figure}[ht]
    \centering
    \includegraphics[width=.6\textwidth]{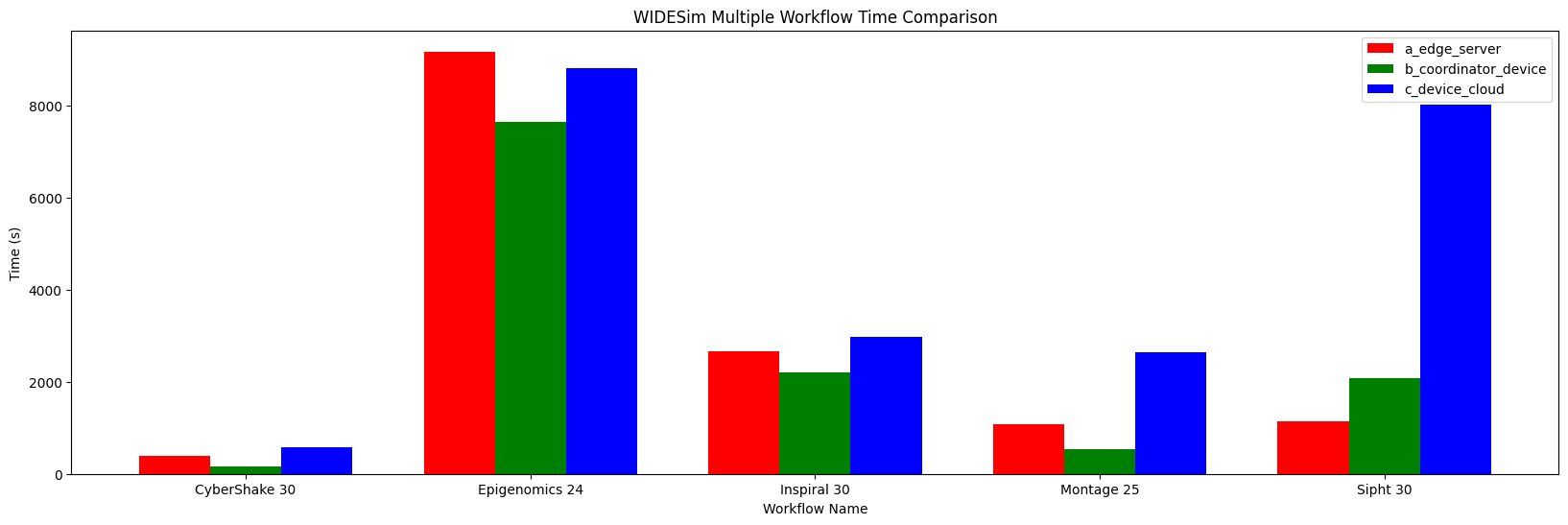}
    \caption{Comparing the execution time of 5 workflows in group execution mode}
    \label{fig:multi1000}
\end{figure}

\begin{figure}[ht]
    \centering
    \includegraphics[width=.6\textwidth]{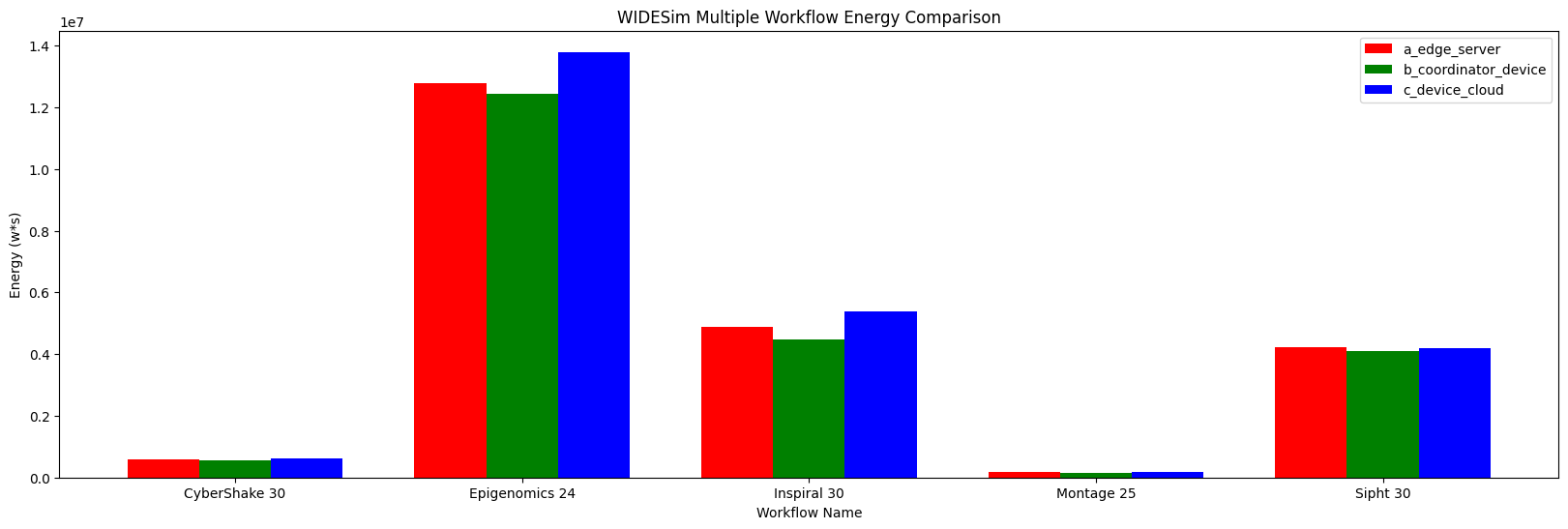}
    \caption{Comparing the energy consumption of 5 workflows in group execution mode}
    \label{fig:multi1001}
\end{figure}

Another issue we examine in this part is the execution time per each workflow. To evaluate this, we examine the start and end times and the duration of each of the workflows in the topologies of section \ref{sec:topologies}. For readability and easier evaluation, we consider a group of five workflows for this issue, which list of them and the results of this study can be seen in table \ref{table:multi_workflow_time_results}. In this table, the start time, end time, and duration for each of the five workflows are reported for each of the three network topologies in the section. As can be seen, all five workflows are being executed in parallel, and this confirms the WIDESim supports evaluation on a group of parallel workflows. Also, comparing this table with the single execution case for FCFS scheduling algorithm \ref{table:compute_results} of each of the workflows shows that most execution times are in harmony with the single case, and in cases where the execution time in the multi-instance is longer than the single workflow case, a few tasks of that workflow have waited in the queue to be scheduled.
For easier review of the amount of time spent by each of the workflows, their execution duration for different topologies is leaked in figure \ref{fig:multi1000}. There is also the comparison of energy consumption in figure \ref{fig:multi1001}.

\begin{table}
\centering
\begin{tabular}{|c|c|c|c|}
 \hline
 \multicolumn{4}{|c |}{Edge Server} \\
 \hline
 \textbf{Workflow} & \textbf{Start Time } & \textbf{End Time } & \textbf{Duration }\\ 
 \hline
 Cybershake 30 & 0.11 & 388.57 & 388.46 \\
 \hline
 Epigenomics 24 & 0.11 & 9175.61 & 9175.5 \\
 \hline
 Inspiral 30 & 0.11 & 2665.74 & 2665.63 \\
 \hline
 Montage 25 & 28.85 & 1107.53 & 1078.68 \\
 \hline
 Sipht 30 & 33.72 & 1185.03 & 1151.31 \\
 \hline
\end{tabular}
\begin{tabular}{|c|c|c|c|}
 \hline
 \multicolumn{4}{|c|}{Coordinator Device} \\
 \hline
 \textbf{Workflow} & \textbf{Start Time } & \textbf{End Time } & \textbf{Duration }\\ 
 \hline
 Cybershake 30 & 0.11 & 159.02 & 159.91 \\
 \hline
 Epigenomics 24 & 0.11 & 7637.38 & 7637.27 \\
 \hline
 Inspiral 30 & 0.11 & 2201.35 & 2201.24 \\
 \hline
 Montage 25 & 12.00 & 556.01 & 544.01 \\
 \hline
 Sipht 30 & 25.99 & 2111.49 & 2085.50 \\
 \hline
\end{tabular}
\begin{tabular}{|c|c|c|c|}
 \hline
 \multicolumn{4}{|c|}{Device Cloud} \\
 \hline
 \textbf{Workflow} & \textbf{Start Time }   & \textbf{End Time } & \textbf{Duration }\\ 
 \hline
 Cybershake 30 & 0.11 & 577.93 & 577.82\\
 \hline
 Epigenomics 24 & 0.11 & 8820.84 & 8820.73\\
 \hline
 Inspiral 30 & 0.11 & 2985.75 & 2985.64 \\
 \hline
 Montage 25 & 36.05 & 2683.93 & 2647.88 \\
 \hline
 Sipht 30 & 47.39 & 8071.13 & 8023.74 \\
 \hline
\end{tabular}
\caption{Investigating the execution time of workflows in group workflow evaluation\label{table:multi_workflow_time_results}}
\end{table}

Referring to Figure~\ref{fig:multi1000}, the execution time for the device cloud topology scenario exceeds that of the other two scenarios for most scientific workflows. This observation aligns with prior findings from single workflow evaluations. The increased execution time in the device cloud scenario arises from the lower computational power and the significant latency due to numerous links to the cloud. Conversely, computational power in this scenario is distributed across multiple devices, ensuring that no single device becomes a system bottleneck as a result of high latency. Furthermore, in all scientific workflows except for Sipht, the coordinator device topology scenario exhibits shorter execution times than the edge server scenario, which can be attributed to its greater computational resources. It is also noteworthy that Sipht, being the last scientific workflow to enter the simulation phase, encounters longer wait times in the queue for processing, as evidenced by Table~\ref{table:multi_workflow_time_results}. This delay, under high system load and due to the non-optimal scheduling algorithm of FCFS, results in increased time intervals between the commencement and completion of tasks in coordinator device topology compared to edge server topology. Additionally, the execution time differential between device cloud scenario and the others notably increases for Montage and Sipht, indicating a higher burden on this topology toward the end of the simulation, attributable to the reduced computational power in the device cloud scenario.

In terms of energy consumption, as illustrated in Figure~\ref{fig:multi1001}, the variance across all three topologies is not as pronounced as it was with execution time. Energy consumption is comparable in most scientific workflows, with the exceptions of Epigenomics and Inspiral. This stands in contrast to the single workflow energy comparisons shown in Figure~\ref{fig:single_workflow_energy_results}, where the coordinator device topology and edge server scenarios exhibited higher power consumption than the device cloud scenario. In the context of multiple workflow simulations, however, the energy consumption of the device cloud topology exceeds that of the other topologies. This increase is due to the accumulation of waiting tasks and the strain of an overloaded system, compounded by the non-optimal scheduling algorithm, which results in higher energy consumption in the device cloud scenario.

\subsubsection{Workflow Ensemble Evaluation}

\begin{figure}[ht]
    \centering
    \includegraphics[width=.8\textwidth]{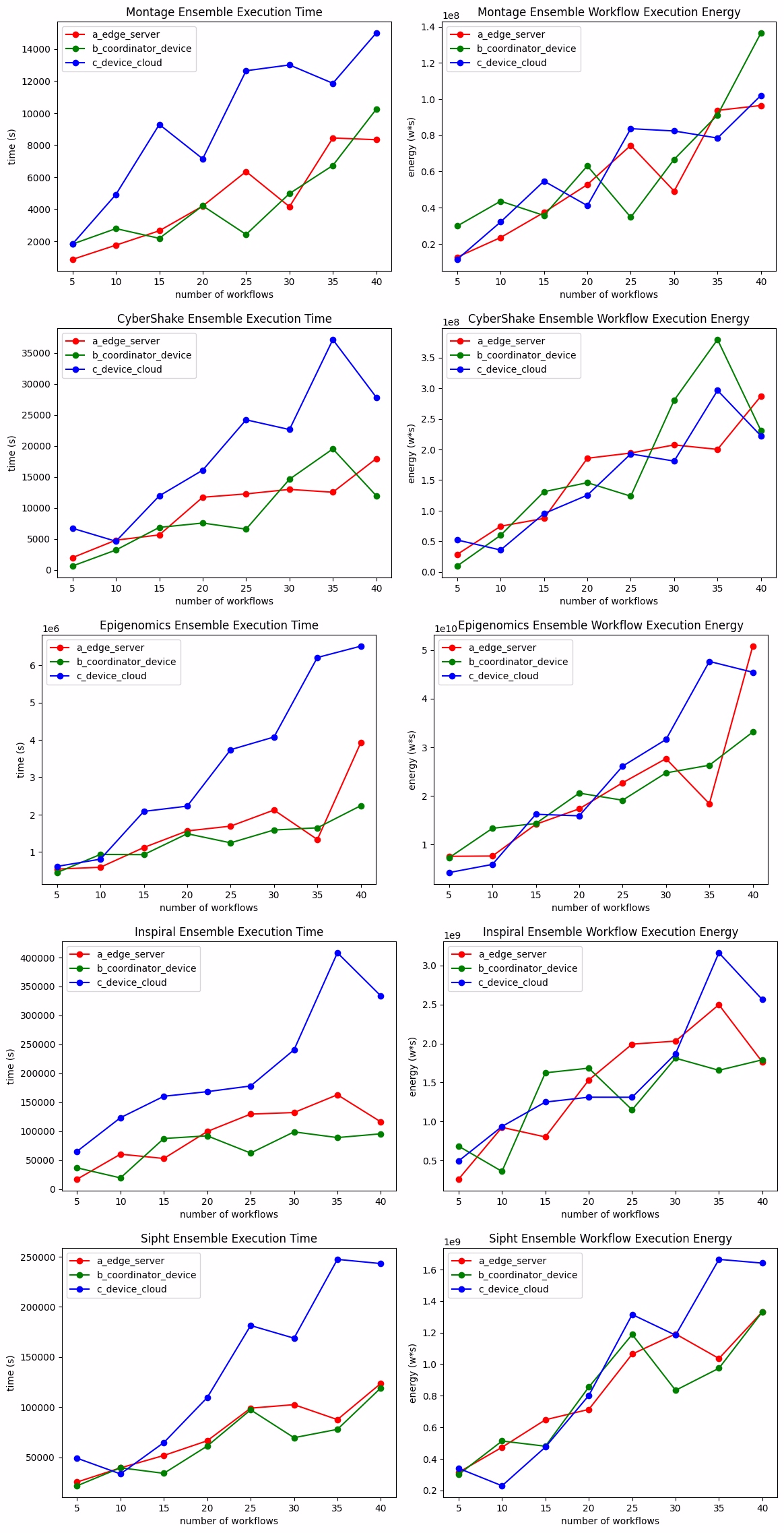}
    \caption{Ensemble workflow evaluation and comparison of time and energy}
    \label{fig:ensemble1}
\end{figure}

In the following, we will evaluate the WIDESim simulator on workflow ensembles that are a group of workflows from one type but with different sizes chosen based on a probability distribution according to the article \cite{malawski2015algorithms}. In this evaluation, for the 5 benchmark topologies, we randomly select some of them with different sizes and concatenate them and then simulate the process. The distribution which is used to pick random sizes of each scientific workflow is uniform. We concatenate and run each of the workflows with a number of 5 through 40 workflows and examine the results. The comparison of these cases in terms of time and energy can be seen in figure \ref{fig:ensemble1}. Also, the implementation code of this simulation is accessible in the \href{https://github.com/ARH80/WIDESim}{\texttt{WIDESim GitHub repository}}. You can refer to the documentation section \href{https://github.com/ARH80/WIDESim/blob/master/doc}{\texttt{doc}} for further information on how to repeat this experiment.

In terms of execution time, we can overlay conclude that the duration of device cloud topology is greater than edge server and edge server is greater than coordinator device, this fact was observed about single workflow and multiple workflow simulation in figure \ref{fig:single_workflow_time_results} and \ref{fig:topology-evalmulti}. In the case of energy consumption comparison, there is again no significant order between three topologies as it was mentioned in multiple workflow evaluation. However, the results are much more varieted that is mainly due to randomness in choosing different sizes of each benchmark workflow. Additionally, contrasting to multiple workflow evaluation, the structure of each workflow in group of workflows in each scenario of ensemble evaluation is the same that, avoiding the impact of a larger workflow in terms of execution time and energy on other workflows.

\section{Conclusion}
\label{conc}   

This paper presents a network simulator on top of CloudSim called WIDESim, which supports different forms of scientific workflows in distributed environments with a graph topology while existing network simulators neither support all types of scientific workflows nor a graph topology for the network. WIDESim enables direct communication between devices supporting device-to-device communication, an important technology in 5G and 6G. So, besides simulating all types of scientific workflows, WIDESim supports decentralized resource management and scheduling, which includes a wide range of schemes in the distributed environments of edge, fog, and cloud computing. We have studied the validity of the network and computational model of WIDESim compared to existing standard simulators. It confirms that WIDESim performs correctly while providing improvements either in the computation or network model. We have also conducted performance studies of WIDESim in resource management and application scheduling for diverse types of workflow-based applications and real-life distributed computing systems including edge-based infrastructures with graph topology. These assessments provide evidence for the suitability of WIDESim in simulating various IoT and distributed computing scenarios, highlighting its exceptional features.

In the future, we aim to expand this study in two key areas. Firstly, we plan to enhance the network model by supporting a broader range of IoT devices, including sensors and actuators. These improvements in the network model also include the support of different messaging protocols that are used widely in IoT like MQTT (Message Queuing Telemetry Transport), AMQP (Advanced Message Queuing Protocol), XMPP (Extensible Messaging and Presence Protocol), and CoAP (Constrained Application Protocol), as they are supported in IoTSim-Edge. Additionally, we will focus on allowing support for device mobility and dynamic changes in device connections within the network. Secondly, in terms of the computational model, we intend to reduce workflow execution overhead by implementing features such as workflow clustering. An implementation of this technique is already available in WorkflowSim, and we aim to provide a more enhanced implementation in WIDESim.

\pagebreak
\newpage

\bibliography{sn-bibliography}


\begin{thebibliography}{49}
\ifx \bisbn   \undefined \def \bisbn  #1{ISBN #1}\fi
\ifx \binits  \undefined \def \binits#1{#1}\fi
\ifx \bauthor  \undefined \def \bauthor#1{#1}\fi
\ifx \batitle  \undefined \def \batitle#1{#1}\fi
\ifx \bjtitle  \undefined \def \bjtitle#1{#1}\fi
\ifx \bvolume  \undefined \def \bvolume#1{\textbf{#1}}\fi
\ifx \byear  \undefined \def \byear#1{#1}\fi
\ifx \bissue  \undefined \def \bissue#1{#1}\fi
\ifx \bfpage  \undefined \def \bfpage#1{#1}\fi
\ifx \blpage  \undefined \def \blpage #1{#1}\fi
\ifx \burl  \undefined \def \burl#1{\textsf{#1}}\fi
\ifx \doiurl  \undefined \def \doiurl#1{\url{https://doi.org/#1}}\fi
\ifx \betal  \undefined \def \betal{\textit{et al.}}\fi
\ifx \binstitute  \undefined \def \binstitute#1{#1}\fi
\ifx \binstitutionaled  \undefined \def \binstitutionaled#1{#1}\fi
\ifx \bctitle  \undefined \def \bctitle#1{#1}\fi
\ifx \beditor  \undefined \def \beditor#1{#1}\fi
\ifx \bpublisher  \undefined \def \bpublisher#1{#1}\fi
\ifx \bbtitle  \undefined \def \bbtitle#1{#1}\fi
\ifx \bedition  \undefined \def \bedition#1{#1}\fi
\ifx \bseriesno  \undefined \def \bseriesno#1{#1}\fi
\ifx \blocation  \undefined \def \blocation#1{#1}\fi
\ifx \bsertitle  \undefined \def \bsertitle#1{#1}\fi
\ifx \bsnm \undefined \def \bsnm#1{#1}\fi
\ifx \bsuffix \undefined \def \bsuffix#1{#1}\fi
\ifx \bparticle \undefined \def \bparticle#1{#1}\fi
\ifx \barticle \undefined \def \barticle#1{#1}\fi
\bibcommenthead
\ifx \bconfdate \undefined \def \bconfdate #1{#1}\fi
\ifx \botherref \undefined \def \botherref #1{#1}\fi
\ifx \url \undefined \def \url#1{\textsf{#1}}\fi
\ifx \bchapter \undefined \def \bchapter#1{#1}\fi
\ifx \bbook \undefined \def \bbook#1{#1}\fi
\ifx \bcomment \undefined \def \bcomment#1{#1}\fi
\ifx \oauthor \undefined \def \oauthor#1{#1}\fi
\ifx \citeauthoryear \undefined \def \citeauthoryear#1{#1}\fi
\ifx \endbibitem  \undefined \def \endbibitem {}\fi
\ifx \bconflocation  \undefined \def \bconflocation#1{#1}\fi
\ifx \arxivurl  \undefined \def \arxivurl#1{\textsf{#1}}\fi
\csname PreBibitemsHook\endcsname

\bibitem[\protect\citeauthoryear{Nardelli et~al.}{2017}]{nardelli2017osmotic}
\begin{barticle}
\bauthor{\bsnm{Nardelli}, \binits{M.}},
\bauthor{\bsnm{Nastic}, \binits{S.}},
\bauthor{\bsnm{Dustdar}, \binits{S.}},
\bauthor{\bsnm{Villari}, \binits{M.}},
\bauthor{\bsnm{Ranjan}, \binits{R.}}:
\batitle{Osmotic flow: Osmotic computing+ iot workflow}.
\bjtitle{IEEE Cloud Computing}
\bvolume{4}(\bissue{2}),
\bfpage{68}--\blpage{75}
(\byear{2017})
\end{barticle}
\endbibitem

\bibitem[\protect\citeauthoryear{Siar and Izadi}{2021}]{siar2021offloading}
\begin{barticle}
\bauthor{\bsnm{Siar}, \binits{H.}},
\bauthor{\bsnm{Izadi}, \binits{M.}}:
\batitle{Offloading coalition formation for scheduling scientific workflow ensembles in fog environments}.
\bjtitle{Journal of Grid Computing}
\bvolume{19}(\bissue{3}),
\bfpage{1}--\blpage{20}
(\byear{2021})
\end{barticle}
\endbibitem

\bibitem[\protect\citeauthoryear{Rodriguez and Buyya}{2018}]{rodriguez2018scheduling}
\begin{barticle}
\bauthor{\bsnm{Rodriguez}, \binits{M.A.}},
\bauthor{\bsnm{Buyya}, \binits{R.}}:
\batitle{Scheduling dynamic workloads in multi-tenant scientific workflow as a service platforms}.
\bjtitle{Future Generation Computer Systems}
\bvolume{79},
\bfpage{739}--\blpage{750}
(\byear{2018})
\end{barticle}
\endbibitem

\bibitem[\protect\citeauthoryear{De~Donno et~al.}{2019}]{de2019foundations}
\begin{barticle}
\bauthor{\bsnm{De~Donno}, \binits{M.}},
\bauthor{\bsnm{Tange}, \binits{K.}},
\bauthor{\bsnm{Dragoni}, \binits{N.}}:
\batitle{Foundations and evolution of modern computing paradigms: Cloud, iot, edge, and fog}.
\bjtitle{Ieee Access}
\bvolume{7},
\bfpage{150936}--\blpage{150948}
(\byear{2019})
\end{barticle}
\endbibitem

\bibitem[\protect\citeauthoryear{Karagiannis et~al.}{2021}]{karagiannis2021context}
\begin{botherref}
\oauthor{\bsnm{Karagiannis}, \binits{V.}},
\oauthor{\bsnm{Frangoudis}, \binits{P.A.}},
\oauthor{\bsnm{Dustdar}, \binits{S.}},
\oauthor{\bsnm{Schulte}, \binits{S.}}:
Context-aware routing in fog computing systems.
IEEE Transactions on Cloud Computing
(2021)
\end{botherref}
\endbibitem

\bibitem[\protect\citeauthoryear{Rabay'a et~al.}{2019}]{rabay2019fog}
\begin{bchapter}
\bauthor{\bsnm{Rabay'a}, \binits{A.}},
\bauthor{\bsnm{Schleicher}, \binits{E.}},
\bauthor{\bsnm{Graffi}, \binits{K.}}:
\bctitle{Fog computing with p2p: Enhancing fog computing bandwidth for iot scenarios}.
In: \bbtitle{2019 International Conference on Internet of Things (iThings) and IEEE Green Computing and Communications (GreenCom) and IEEE Cyber, Physical and Social Computing (CPSCom) and IEEE Smart Data (SmartData)},
pp. \bfpage{82}--\blpage{89}
(\byear{2019}).
\bcomment{IEEE}
\end{bchapter}
\endbibitem

\bibitem[\protect\citeauthoryear{Yousefpour et~al.}{2019}]{yousefpour2019all}
\begin{barticle}
\bauthor{\bsnm{Yousefpour}, \binits{A.}},
\bauthor{\bsnm{Fung}, \binits{C.}},
\bauthor{\bsnm{Nguyen}, \binits{T.}},
\bauthor{\bsnm{Kadiyala}, \binits{K.}},
\bauthor{\bsnm{Jalali}, \binits{F.}},
\bauthor{\bsnm{Niakanlahiji}, \binits{A.}},
\bauthor{\bsnm{Kong}, \binits{J.}},
\bauthor{\bsnm{Jue}, \binits{J.P.}}:
\batitle{All one needs to know about fog computing and related edge computing paradigms: A complete survey}.
\bjtitle{Journal of Systems Architecture}
\bvolume{98},
\bfpage{289}--\blpage{330}
(\byear{2019})
\end{barticle}
\endbibitem

\bibitem[\protect\citeauthoryear{Tocz{\'e} and Nadjm-Tehrani}{2018}]{tocze2018taxonomy}
\begin{botherref}
\oauthor{\bsnm{Tocz{\'e}}, \binits{K.}},
\oauthor{\bsnm{Nadjm-Tehrani}, \binits{S.}}:
A taxonomy for management and optimization of multiple resources in edge computing.
Wireless Communications and Mobile Computing
\textbf{2018}
(2018)
\end{botherref}
\endbibitem

\bibitem[\protect\citeauthoryear{Buyya and Srirama}{2019}]{buyya2019fog}
\begin{bbook}
\bauthor{\bsnm{Buyya}, \binits{R.}},
\bauthor{\bsnm{Srirama}, \binits{S.N.}}:
\bbtitle{Fog and Edge Computing: Principles and Paradigms}.
\bpublisher{John Wiley \& Sons}, \blocation{???}
(\byear{2019})
\end{bbook}
\endbibitem

\bibitem[\protect\citeauthoryear{Yu et~al.}{2015}]{yu2015cooperative}
\begin{barticle}
\bauthor{\bsnm{Yu}, \binits{R.}},
\bauthor{\bsnm{Huang}, \binits{X.}},
\bauthor{\bsnm{Kang}, \binits{J.}},
\bauthor{\bsnm{Ding}, \binits{J.}},
\bauthor{\bsnm{Maharjan}, \binits{S.}},
\bauthor{\bsnm{Gjessing}, \binits{S.}},
\bauthor{\bsnm{Zhang}, \binits{Y.}}:
\batitle{Cooperative resource management in cloud-enabled vehicular networks}.
\bjtitle{IEEE Transactions on Industrial Electronics}
\bvolume{62}(\bissue{12}),
\bfpage{7938}--\blpage{7951}
(\byear{2015})
\end{barticle}
\endbibitem

\bibitem[\protect\citeauthoryear{Stavrinides and Karatza}{2019}]{stavrinides2019hybrid}
\begin{barticle}
\bauthor{\bsnm{Stavrinides}, \binits{G.L.}},
\bauthor{\bsnm{Karatza}, \binits{H.D.}}:
\batitle{A hybrid approach to scheduling real-time iot workflows in fog and cloud environments}.
\bjtitle{Multimedia Tools and Applications}
\bvolume{78}(\bissue{17}),
\bfpage{24639}--\blpage{24655}
(\byear{2019})
\end{barticle}
\endbibitem

\bibitem[\protect\citeauthoryear{Chen and Xu}{2017}]{chen2017socially}
\begin{bchapter}
\bauthor{\bsnm{Chen}, \binits{L.}},
\bauthor{\bsnm{Xu}, \binits{J.}}:
\bctitle{Socially trusted collaborative edge computing in ultra dense networks}.
In: \bbtitle{Proceedings of the Second ACM/IEEE Symposium on Edge Computing},
pp. \bfpage{1}--\blpage{11}
(\byear{2017})
\end{bchapter}
\endbibitem

\bibitem[\protect\citeauthoryear{Lim et~al.}{2020}]{lim2020incentive}
\begin{botherref}
\oauthor{\bsnm{Lim}, \binits{W.Y.B.}},
\oauthor{\bsnm{Ng}, \binits{J.S.}},
\oauthor{\bsnm{Xiong}, \binits{Z.}},
\oauthor{\bsnm{Niyato}, \binits{D.}},
\oauthor{\bsnm{Leung}, \binits{C.}},
\oauthor{\bsnm{Miao}, \binits{C.}},
\oauthor{\bsnm{Yang}, \binits{Q.}}:
Incentive mechanism design for resource sharing in collaborative edge learning.
arXiv preprint arXiv:2006.00511
(2020)
\end{botherref}
\endbibitem

\bibitem[\protect\citeauthoryear{Bianzino et~al.}{2014}]{bianzino2014green}
\begin{barticle}
\bauthor{\bsnm{Bianzino}, \binits{A.P.}},
\bauthor{\bsnm{Rougier}, \binits{J.-L.}},
\bauthor{\bsnm{Chaudet}, \binits{C.}},
\bauthor{\bsnm{Rossi}, \binits{D.}}, \betal:
\batitle{The green-game: Accounting for device criticality in resource consolidation for backbone ip networks}.
\bjtitle{Strategic Behavior and the Environment}
\bvolume{4}(\bissue{2}),
\bfpage{131}--\blpage{153}
(\byear{2014})
\end{barticle}
\endbibitem

\bibitem[\protect\citeauthoryear{Goudarzi et~al.}{2020}]{goudarzi2020application}
\begin{barticle}
\bauthor{\bsnm{Goudarzi}, \binits{M.}},
\bauthor{\bsnm{Wu}, \binits{H.}},
\bauthor{\bsnm{Palaniswami}, \binits{M.}},
\bauthor{\bsnm{Buyya}, \binits{R.}}:
\batitle{An application placement technique for concurrent iot applications in edge and fog computing environments}.
\bjtitle{IEEE Transactions on Mobile Computing}
\bvolume{20}(\bissue{4}),
\bfpage{1298}--\blpage{1311}
(\byear{2020})
\end{barticle}
\endbibitem

\bibitem[\protect\citeauthoryear{Zhang et~al.}{2019}]{zhang2019dmra}
\begin{bchapter}
\bauthor{\bsnm{Zhang}, \binits{C.}},
\bauthor{\bsnm{Du}, \binits{H.}},
\bauthor{\bsnm{Ye}, \binits{Q.}},
\bauthor{\bsnm{Liu}, \binits{C.}},
\bauthor{\bsnm{Yuan}, \binits{H.}}:
\bctitle{Dmra: a decentralized resource allocation scheme for multi-sp mobile edge computing}.
In: \bbtitle{2019 IEEE 39th International Conference on Distributed Computing Systems (ICDCS)},
pp. \bfpage{390}--\blpage{398}
(\byear{2019}).
\bcomment{IEEE}
\end{bchapter}
\endbibitem

\bibitem[\protect\citeauthoryear{Tripathi et~al.}{2017}]{tripathi2017non}
\begin{barticle}
\bauthor{\bsnm{Tripathi}, \binits{R.}},
\bauthor{\bsnm{Vignesh}, \binits{S.}},
\bauthor{\bsnm{Tamarapalli}, \binits{V.}},
\bauthor{\bsnm{Chronopoulos}, \binits{A.T.}},
\bauthor{\bsnm{Siar}, \binits{H.}}:
\batitle{Non-cooperative power and latency aware load balancing in distributed data centers}.
\bjtitle{Journal of Parallel and Distributed Computing}
\bvolume{107},
\bfpage{76}--\blpage{86}
(\byear{2017})
\end{barticle}
\endbibitem

\bibitem[\protect\citeauthoryear{Jo{\v{s}}ilo and D{\'a}n}{2018}]{jovsilo2018decentralized}
\begin{barticle}
\bauthor{\bsnm{Jo{\v{s}}ilo}, \binits{S.}},
\bauthor{\bsnm{D{\'a}n}, \binits{G.}}:
\batitle{Decentralized algorithm for randomized task allocation in fog computing systems}.
\bjtitle{IEEE/ACM Transactions on Networking}
\bvolume{27}(\bissue{1}),
\bfpage{85}--\blpage{97}
(\byear{2018})
\end{barticle}
\endbibitem

\bibitem[\protect\citeauthoryear{Guerrero et~al.}{2019}]{guerrero2019lightweight}
\begin{barticle}
\bauthor{\bsnm{Guerrero}, \binits{C.}},
\bauthor{\bsnm{Lera}, \binits{I.}},
\bauthor{\bsnm{Juiz}, \binits{C.}}:
\batitle{A lightweight decentralized service placement policy for performance optimization in fog computing}.
\bjtitle{Journal of Ambient Intelligence and Humanized Computing}
\bvolume{10}(\bissue{6}),
\bfpage{2435}--\blpage{2452}
(\byear{2019})
\end{barticle}
\endbibitem

\bibitem[\protect\citeauthoryear{Gupta et~al.}{2017}]{gupta2017ifogsim}
\begin{barticle}
\bauthor{\bsnm{Gupta}, \binits{H.}},
\bauthor{\bsnm{Vahid~Dastjerdi}, \binits{A.}},
\bauthor{\bsnm{Ghosh}, \binits{S.K.}},
\bauthor{\bsnm{Buyya}, \binits{R.}}:
\batitle{ifogsim: A toolkit for modeling and simulation of resource management techniques in the internet of things, edge and fog computing environments}.
\bjtitle{Software: Practice and Experience}
\bvolume{47}(\bissue{9}),
\bfpage{1275}--\blpage{1296}
(\byear{2017})
\end{barticle}
\endbibitem

\bibitem[\protect\citeauthoryear{Chen and Deelman}{2012}]{chen2012workflowsim}
\begin{bchapter}
\bauthor{\bsnm{Chen}, \binits{W.}},
\bauthor{\bsnm{Deelman}, \binits{E.}}:
\bctitle{Workflowsim: A toolkit for simulating scientific workflows in distributed environments}.
In: \bbtitle{2012 IEEE 8th International Conference on E-science},
pp. \bfpage{1}--\blpage{8}
(\byear{2012}).
\bcomment{IEEE}
\end{bchapter}
\endbibitem

\bibitem[\protect\citeauthoryear{Liu et~al.}{2019}]{liu2019fogworkflowsim}
\begin{bchapter}
\bauthor{\bsnm{Liu}, \binits{X.}},
\bauthor{\bsnm{Fan}, \binits{L.}},
\bauthor{\bsnm{Xu}, \binits{J.}},
\bauthor{\bsnm{Li}, \binits{X.}},
\bauthor{\bsnm{Gong}, \binits{L.}},
\bauthor{\bsnm{Grundy}, \binits{J.}},
\bauthor{\bsnm{Yang}, \binits{Y.}}:
\bctitle{Fogworkflowsim: An automated simulation toolkit for workflow performance evaluation in fog computing}.
In: \bbtitle{2019 34th IEEE/ACM International Conference on Automated Software Engineering (ASE)},
pp. \bfpage{1114}--\blpage{1117}
(\byear{2019}).
\bcomment{IEEE}
\end{bchapter}
\endbibitem

\bibitem[\protect\citeauthoryear{Calheiros et~al.}{2011}]{calheiros2011cloudsim}
\begin{barticle}
\bauthor{\bsnm{Calheiros}, \binits{R.N.}},
\bauthor{\bsnm{Ranjan}, \binits{R.}},
\bauthor{\bsnm{Beloglazov}, \binits{A.}},
\bauthor{\bsnm{De~Rose}, \binits{C.A.}},
\bauthor{\bsnm{Buyya}, \binits{R.}}:
\batitle{Cloudsim: a toolkit for modeling and simulation of cloud computing environments and evaluation of resource provisioning algorithms}.
\bjtitle{Software: Practice and experience}
\bvolume{41}(\bissue{1}),
\bfpage{23}--\blpage{50}
(\byear{2011})
\end{barticle}
\endbibitem

\bibitem[\protect\citeauthoryear{Hameed et~al.}{2016}]{hameed2016survey}
\begin{barticle}
\bauthor{\bsnm{Hameed}, \binits{A.}},
\bauthor{\bsnm{Khoshkbarforoushha}, \binits{A.}},
\bauthor{\bsnm{Ranjan}, \binits{R.}},
\bauthor{\bsnm{Jayaraman}, \binits{P.P.}},
\bauthor{\bsnm{Kolodziej}, \binits{J.}},
\bauthor{\bsnm{Balaji}, \binits{P.}},
\bauthor{\bsnm{Zeadally}, \binits{S.}},
\bauthor{\bsnm{Malluhi}, \binits{Q.M.}},
\bauthor{\bsnm{Tziritas}, \binits{N.}},
\bauthor{\bsnm{Vishnu}, \binits{A.}}, \betal:
\batitle{A survey and taxonomy on energy efficient resource allocation techniques for cloud computing systems}.
\bjtitle{Computing}
\bvolume{98}(\bissue{7}),
\bfpage{751}--\blpage{774}
(\byear{2016})
\end{barticle}
\endbibitem

\bibitem[\protect\citeauthoryear{Zhang et~al.}{2010}]{zhang2010cloud}
\begin{barticle}
\bauthor{\bsnm{Zhang}, \binits{Q.}},
\bauthor{\bsnm{Cheng}, \binits{L.}},
\bauthor{\bsnm{Boutaba}, \binits{R.}}:
\batitle{Cloud computing: state-of-the-art and research challenges}.
\bjtitle{Journal of internet services and applications}
\bvolume{1}(\bissue{1}),
\bfpage{7}--\blpage{18}
(\byear{2010})
\end{barticle}
\endbibitem

\bibitem[\protect\citeauthoryear{Singh and Chana}{2016}]{singh2016survey}
\begin{barticle}
\bauthor{\bsnm{Singh}, \binits{S.}},
\bauthor{\bsnm{Chana}, \binits{I.}}:
\batitle{A survey on resource scheduling in cloud computing: Issues and challenges}.
\bjtitle{Journal of grid computing}
\bvolume{14}(\bissue{2}),
\bfpage{217}--\blpage{264}
(\byear{2016})
\end{barticle}
\endbibitem

\bibitem[\protect\citeauthoryear{Yi et~al.}{2015}]{yi2015survey}
\begin{bchapter}
\bauthor{\bsnm{Yi}, \binits{S.}},
\bauthor{\bsnm{Li}, \binits{C.}},
\bauthor{\bsnm{Li}, \binits{Q.}}:
\bctitle{A survey of fog computing: concepts, applications and issues}.
In: \bbtitle{Proceedings of the 2015 Workshop on Mobile Big Data},
pp. \bfpage{37}--\blpage{42}
(\byear{2015})
\end{bchapter}
\endbibitem

\bibitem[\protect\citeauthoryear{Zhang et~al.}{2017}]{zhang2017hierarchical}
\begin{barticle}
\bauthor{\bsnm{Zhang}, \binits{H.}},
\bauthor{\bsnm{Zhang}, \binits{Y.}},
\bauthor{\bsnm{Gu}, \binits{Y.}},
\bauthor{\bsnm{Niyato}, \binits{D.}},
\bauthor{\bsnm{Han}, \binits{Z.}}:
\batitle{A hierarchical game framework for resource management in fog computing}.
\bjtitle{IEEE Communications Magazine}
\bvolume{55}(\bissue{8}),
\bfpage{52}--\blpage{57}
(\byear{2017})
\end{barticle}
\endbibitem

\bibitem[\protect\citeauthoryear{Mahmud et~al.}{2020}]{mahmud2020application}
\begin{barticle}
\bauthor{\bsnm{Mahmud}, \binits{R.}},
\bauthor{\bsnm{Ramamohanarao}, \binits{K.}},
\bauthor{\bsnm{Buyya}, \binits{R.}}:
\batitle{Application management in fog computing environments: A taxonomy, review and future directions}.
\bjtitle{ACM Computing Surveys (CSUR)}
\bvolume{53}(\bissue{4}),
\bfpage{1}--\blpage{43}
(\byear{2020})
\end{barticle}
\endbibitem

\bibitem[\protect\citeauthoryear{Xu et~al.}{2021}]{xu2021privacy}
\begin{barticle}
\bauthor{\bsnm{Xu}, \binits{H.}},
\bauthor{\bsnm{Qiu}, \binits{X.}},
\bauthor{\bsnm{Zhang}, \binits{W.}},
\bauthor{\bsnm{Liu}, \binits{K.}},
\bauthor{\bsnm{Liu}, \binits{S.}},
\bauthor{\bsnm{Chen}, \binits{W.}}:
\batitle{Privacy-preserving incentive mechanism for multi-leader multi-follower iot-edge computing market: A reinforcement learning approach}.
\bjtitle{Journal of Systems Architecture}
\bvolume{114},
\bfpage{101932}
(\byear{2021})
\end{barticle}
\endbibitem

\bibitem[\protect\citeauthoryear{Li et~al.}{2020}]{li2020noma}
\begin{barticle}
\bauthor{\bsnm{Li}, \binits{Z.}},
\bauthor{\bsnm{Xu}, \binits{M.}},
\bauthor{\bsnm{Nie}, \binits{J.}},
\bauthor{\bsnm{Kang}, \binits{J.}},
\bauthor{\bsnm{Chen}, \binits{W.}},
\bauthor{\bsnm{Xie}, \binits{S.}}:
\batitle{Noma-enabled cooperative computation offloading for blockchain-empowered internet of things: A learning approach}.
\bjtitle{IEEE Internet of Things Journal}
\bvolume{8}(\bissue{4}),
\bfpage{2364}--\blpage{2378}
(\byear{2020})
\end{barticle}
\endbibitem

\bibitem[\protect\citeauthoryear{Ijaz et~al.}{2021}]{ijaz2021energy}
\begin{barticle}
\bauthor{\bsnm{Ijaz}, \binits{S.}},
\bauthor{\bsnm{Munir}, \binits{E.U.}},
\bauthor{\bsnm{Ahmad}, \binits{S.G.}},
\bauthor{\bsnm{Rafique}, \binits{M.M.}},
\bauthor{\bsnm{Rana}, \binits{O.F.}}:
\batitle{Energy-makespan optimization of workflow scheduling in fog--cloud computing}.
\bjtitle{Computing}
\bvolume{103}(\bissue{9}),
\bfpage{2033}--\blpage{2059}
(\byear{2021})
\end{barticle}
\endbibitem

\bibitem[\protect\citeauthoryear{Hong et~al.}{2019}]{hong2019multi}
\begin{barticle}
\bauthor{\bsnm{Hong}, \binits{Z.}},
\bauthor{\bsnm{Chen}, \binits{W.}},
\bauthor{\bsnm{Huang}, \binits{H.}},
\bauthor{\bsnm{Guo}, \binits{S.}},
\bauthor{\bsnm{Zheng}, \binits{Z.}}:
\batitle{Multi-hop cooperative computation offloading for industrial iot--edge--cloud computing environments}.
\bjtitle{IEEE Transactions on Parallel and Distributed Systems}
\bvolume{30}(\bissue{12}),
\bfpage{2759}--\blpage{2774}
(\byear{2019})
\end{barticle}
\endbibitem

\bibitem[\protect\citeauthoryear{Zeng et~al.}{2016}]{zeng2016joint}
\begin{barticle}
\bauthor{\bsnm{Zeng}, \binits{D.}},
\bauthor{\bsnm{Gu}, \binits{L.}},
\bauthor{\bsnm{Guo}, \binits{S.}},
\bauthor{\bsnm{Cheng}, \binits{Z.}},
\bauthor{\bsnm{Yu}, \binits{S.}}:
\batitle{Joint optimization of task scheduling and image placement in fog computing supported software-defined embedded system}.
\bjtitle{IEEE Transactions on Computers}
\bvolume{65}(\bissue{12}),
\bfpage{3702}--\blpage{3712}
(\byear{2016})
\end{barticle}
\endbibitem

\bibitem[\protect\citeauthoryear{Mahmud et~al.}{2018}]{mahmud2018latency}
\begin{barticle}
\bauthor{\bsnm{Mahmud}, \binits{R.}},
\bauthor{\bsnm{Ramamohanarao}, \binits{K.}},
\bauthor{\bsnm{Buyya}, \binits{R.}}:
\batitle{Latency-aware application module management for fog computing environments}.
\bjtitle{ACM Transactions on Internet Technology (TOIT)}
\bvolume{19}(\bissue{1}),
\bfpage{1}--\blpage{21}
(\byear{2018})
\end{barticle}
\endbibitem

\bibitem[\protect\citeauthoryear{Xu et~al.}{2019}]{xu2019computation}
\begin{barticle}
\bauthor{\bsnm{Xu}, \binits{X.}},
\bauthor{\bsnm{Liu}, \binits{Q.}},
\bauthor{\bsnm{Luo}, \binits{Y.}},
\bauthor{\bsnm{Peng}, \binits{K.}},
\bauthor{\bsnm{Zhang}, \binits{X.}},
\bauthor{\bsnm{Meng}, \binits{S.}},
\bauthor{\bsnm{Qi}, \binits{L.}}:
\batitle{A computation offloading method over big data for iot-enabled cloud-edge computing}.
\bjtitle{Future Generation Computer Systems}
\bvolume{95},
\bfpage{522}--\blpage{533}
(\byear{2019})
\end{barticle}
\endbibitem

\bibitem[\protect\citeauthoryear{de~Souza~Toniolli and Jaumard}{2019}]{de2019resource}
\begin{bchapter}
\bauthor{\bsnm{Souza~Toniolli}, \binits{J.L.}},
\bauthor{\bsnm{Jaumard}, \binits{B.}}:
\bctitle{Resource allocation for multiple workflows in cloud-fog computing systems}.
In: \bbtitle{Proceedings of the 12th IEEE/ACM International Conference on Utility and Cloud Computing Companion},
pp. \bfpage{77}--\blpage{84}
(\byear{2019})
\end{bchapter}
\endbibitem

\bibitem[\protect\citeauthoryear{De~Maio and Kimovski}{2020}]{de2020multi}
\begin{barticle}
\bauthor{\bsnm{De~Maio}, \binits{V.}},
\bauthor{\bsnm{Kimovski}, \binits{D.}}:
\batitle{Multi-objective scheduling of extreme data scientific workflows in fog}.
\bjtitle{Future Generation Computer Systems}
\bvolume{106},
\bfpage{171}--\blpage{184}
(\byear{2020})
\end{barticle}
\endbibitem

\bibitem[\protect\citeauthoryear{Bharathi et~al.}{2008}]{bharathi2008characterization}
\begin{bchapter}
\bauthor{\bsnm{Bharathi}, \binits{S.}},
\bauthor{\bsnm{Chervenak}, \binits{A.}},
\bauthor{\bsnm{Deelman}, \binits{E.}},
\bauthor{\bsnm{Mehta}, \binits{G.}},
\bauthor{\bsnm{Su}, \binits{M.-H.}},
\bauthor{\bsnm{Vahi}, \binits{K.}}:
\bctitle{Characterization of scientific workflows}.
In: \bbtitle{2008 Third Workshop on Workflows in Support of Large-scale Science},
pp. \bfpage{1}--\blpage{10}
(\byear{2008}).
\bcomment{IEEE}
\end{bchapter}
\endbibitem

\bibitem[\protect\citeauthoryear{Genez et~al.}{2017}]{genez2017robust}
\begin{bchapter}
\bauthor{\bsnm{Genez}, \binits{T.A.L.}},
\bauthor{\bsnm{Bittencourt}, \binits{L.F.}},
\bauthor{\bsnm{Sakellariou}, \binits{R.}},
\bauthor{\bsnm{Madeira}, \binits{E.R.M.}}:
\bctitle{A robust scheduler for workflow ensembles under uncertainties of available bandwidth}.
In: \bbtitle{2017 IEEE 10th International Conference on Cloud Computing (CLOUD)},
pp. \bfpage{504}--\blpage{511}
(\byear{2017}).
\bcomment{IEEE}
\end{bchapter}
\endbibitem

\bibitem[\protect\citeauthoryear{Jiang et~al.}{2015}]{jiang2015executing}
\begin{bchapter}
\bauthor{\bsnm{Jiang}, \binits{Q.}},
\bauthor{\bsnm{Lee}, \binits{Y.C.}},
\bauthor{\bsnm{Zomaya}, \binits{A.Y.}}:
\bctitle{Executing large scale scientific workflow ensembles in public clouds}.
In: \bbtitle{2015 44th International Conference on Parallel Processing},
pp. \bfpage{520}--\blpage{529}
(\byear{2015}).
\bcomment{IEEE}
\end{bchapter}
\endbibitem

\bibitem[\protect\citeauthoryear{Malawski et~al.}{2015}]{malawski2015algorithms}
\begin{barticle}
\bauthor{\bsnm{Malawski}, \binits{M.}},
\bauthor{\bsnm{Juve}, \binits{G.}},
\bauthor{\bsnm{Deelman}, \binits{E.}},
\bauthor{\bsnm{Nabrzyski}, \binits{J.}}:
\batitle{Algorithms for cost-and deadline-constrained provisioning for scientific workflow ensembles in iaas clouds}.
\bjtitle{Future Generation Computer Systems}
\bvolume{48},
\bfpage{1}--\blpage{18}
(\byear{2015})
\end{barticle}
\endbibitem

\bibitem[\protect\citeauthoryear{Taylor et~al.}{2007}]{taylor2007workflows}
\begin{bbook}
\bauthor{\bsnm{Taylor}, \binits{I.J.}},
\bauthor{\bsnm{Deelman}, \binits{E.}},
\bauthor{\bsnm{Gannon}, \binits{D.B.}},
\bauthor{\bsnm{Shields}, \binits{M.}}, \betal:
\bbtitle{Workflows for e-Science: Scientific Workflows for Grids}
vol. \bseriesno{1}.
\bpublisher{Springer}, \blocation{???}
(\byear{2007})
\end{bbook}
\endbibitem

\bibitem[\protect\citeauthoryear{Genez et~al.}{2016}]{genez2016flexible}
\begin{bchapter}
\bauthor{\bsnm{Genez}, \binits{T.A.}},
\bauthor{\bsnm{Bittencourt}, \binits{L.F.}},
\bauthor{\bsnm{Sakellariou}, \binits{R.}},
\bauthor{\bsnm{Madeira}, \binits{E.R.}}:
\bctitle{A flexible scheduler for workflow ensembles}.
In: \bbtitle{Proceedings of the 9th International Conference on Utility and Cloud Computing},
pp. \bfpage{55}--\blpage{62}
(\byear{2016})
\end{bchapter}
\endbibitem

\bibitem[\protect\citeauthoryear{Pietri et~al.}{2013}]{pietri2013energy}
\begin{bchapter}
\bauthor{\bsnm{Pietri}, \binits{I.}},
\bauthor{\bsnm{Malawski}, \binits{M.}},
\bauthor{\bsnm{Juve}, \binits{G.}},
\bauthor{\bsnm{Deelman}, \binits{E.}},
\bauthor{\bsnm{Nabrzyski}, \binits{J.}},
\bauthor{\bsnm{Sakellariou}, \binits{R.}}:
\bctitle{Energy-constrained provisioning for scientific workflow ensembles}.
In: \bbtitle{2013 International Conference on Cloud and Green Computing},
pp. \bfpage{34}--\blpage{41}
(\byear{2013}).
\bcomment{IEEE}
\end{bchapter}
\endbibitem

\bibitem[\protect\citeauthoryear{Mahmud et~al.}{2022}]{mahmud2022ifogsim2}
\begin{barticle}
\bauthor{\bsnm{Mahmud}, \binits{R.}},
\bauthor{\bsnm{Pallewatta}, \binits{S.}},
\bauthor{\bsnm{Goudarzi}, \binits{M.}},
\bauthor{\bsnm{Buyya}, \binits{R.}}:
\batitle{ifogsim2: An extended ifogsim simulator for mobility, clustering, and microservice management in edge and fog computing environments}.
\bjtitle{Journal of Systems and Software}
\bvolume{190},
\bfpage{111351}
(\byear{2022})
\end{barticle}
\endbibitem

\bibitem[\protect\citeauthoryear{Jha et~al.}{2020}]{jha2020iotsim}
\begin{barticle}
\bauthor{\bsnm{Jha}, \binits{D.N.}},
\bauthor{\bsnm{Alwasel}, \binits{K.}},
\bauthor{\bsnm{Alshoshan}, \binits{A.}},
\bauthor{\bsnm{Huang}, \binits{X.}},
\bauthor{\bsnm{Naha}, \binits{R.K.}},
\bauthor{\bsnm{Battula}, \binits{S.K.}},
\bauthor{\bsnm{Garg}, \binits{S.}},
\bauthor{\bsnm{Puthal}, \binits{D.}},
\bauthor{\bsnm{James}, \binits{P.}},
\bauthor{\bsnm{Zomaya}, \binits{A.}}, \betal:
\batitle{Iotsim-edge: a simulation framework for modeling the behavior of internet of things and edge computing environments}.
\bjtitle{Software: Practice and Experience}
\bvolume{50}(\bissue{6}),
\bfpage{844}--\blpage{867}
(\byear{2020})
\end{barticle}
\endbibitem

\bibitem[\protect\citeauthoryear{Sonmez et~al.}{2017}]{sonmez2017EdgeCloudSim}
\begin{botherref}
\oauthor{\bsnm{Sonmez}, \binits{C.}},
\oauthor{\bsnm{Ozgovde}, \binits{A.}},
\oauthor{\bsnm{Ersoy}, \binits{C.}}:
Edgecloudsim: An environment for performance evaluation of edge computing systems,
39--44
(2017)
\end{botherref}
\endbibitem

\bibitem[\protect\citeauthoryear{Juve et~al.}{2013}]{juve2013characterizing}
\begin{barticle}
\bauthor{\bsnm{Juve}, \binits{G.}},
\bauthor{\bsnm{Chervenak}, \binits{A.}},
\bauthor{\bsnm{Deelman}, \binits{E.}},
\bauthor{\bsnm{Bharathi}, \binits{S.}},
\bauthor{\bsnm{Mehta}, \binits{G.}},
\bauthor{\bsnm{Vahi}, \binits{K.}}:
\batitle{Characterizing and profiling scientific workflows}.
\bjtitle{Future generation computer systems}
\bvolume{29}(\bissue{3}),
\bfpage{682}--\blpage{692}
(\byear{2013})
\end{barticle}
\endbibitem

\end{thebibliography}

\end{document}